\documentclass[aps,twocolumn,preprintnumbers,amsmath,amssymb,floatfix,nofootinbib,groupedaddress,nobibnotes]{revtex4-1}
\usepackage{amsmath,amssymb,latexsym,amsthm,ulem}
\usepackage{verbatim}
\usepackage{wrapfig}
\usepackage{times}
\usepackage{epsfig}
\usepackage{hyperref}
\usepackage{graphicx}
\usepackage{color}
\usepackage{dcolumn}
\usepackage{bm}
\usepackage{setspace}
\usepackage{epstopdf}
\usepackage[loose,nice]{units} 
\usepackage{subcaption}
\usepackage{ragged2e}
\DeclareCaptionJustification{justified}{\justifying}
\usepackage{MnSymbol}
\definecolor{orange}{rgb}{1.0,0.4,0.0}
\usepackage{float}	
\usepackage{lipsum}
\usepackage{footnote}
\usepackage{titlesec}
\usepackage{tikz}
\usetikzlibrary{plotmarks}
\usepackage[none]{hyphenat}
\usepackage{hyperref}

\newcommand\marksymbol[2]{\tikz[#2,scale=1.4]\pgfuseplotmark{#1};}
\newcommand\solidrule[1][0.5cm]{\rule[0.5ex]{#1}{.4pt}}
	
	\newcommand\dashedrule{\mbox{%
    	\solidrule[2mm]\hspace{2mm}\solidrule[2mm]\hspace{2mm}\solidrule[2mm]}}
	
	\titlespacing\paragraph{0pt}{12pt plus 4pt minus 2pt}{0pt plus 2pt minus 2pt}
	\newcommand{\bigO}{\ensuremath{\mathcal{O}}}
\newcommand \E[1] {Eq.~(\ref{#1})}

\begin{document}

\title{Analysis of Thermo-Diffusive Cellular Instabilities in Continuum Combustion Fronts}

\author{Hossein Azizi}
\author{Sebastian Gurevich}
\author{Nikolas Provatas}

\affiliation{Department of Physics, Centre for the Physics of Materials, McGill University, Montreal, QC, Canada\\ }

\date{\today}

\begin{abstract}
We explore numerically the morphological patterns of thermo-diffusive instabilities in combustion fronts with a continuum fuel source, within a range of Lewis numbers and ignition temperatures, focusing on the cellular regime. For this purpose, we generalize the model of Brailovsky et al. to include distinct process kinetics and reactant heterogeneity. The generalized model is derived analytically and validated with other established models in the limit of infinite Lewis number for zero-order and first-order kinetics. Cellular and dendritic instabilities are found at low Lewis numbers thanks to a dynamic adaptive mesh refinement technique that reduces finite size effects, which can affect or even preclude the emergence of these patterns. This technique also allows achieving very large computational domains, enabling the study of system-size effects. Our numerical linear stability analysis is consistent with the analytical results of Brailovsky et al.
The distinct types of dynamics found in the vicinity of the critical Lewis number, ranging from steady-state cells to continued tip-splitting and cell-merging, are well described within the framework of thermo-diffusive instabilities and are consistent with previous numerical studies. These types of dynamics are classified as  ``quasi-linear'' and characterized by low amplitude cells and may follow the mode selection mechanism and growth prescribed by the linear theory. Below this range of Lewis number, highly non-linear effects become prominent and large amplitude, complex cellular and {\it{seaweed}} dendritic morphologies emerge.
 \end{abstract}
 
\maketitle

\section{Introduction}
\label{Intro}
Combustion is a process of fast oxidation involving fuel and an oxidant. It is an exothermic chemical reaction with substantial heat release, which through the competition between heat dissipation and mass transport can form a self sustaining propagating front.
A combustion phenomenon of particular interest is that of solid-gas combustion, where solid fuel particle distributions react in the presence of a gaseous oxidizer in which they are dispersed. 

Combustion is also an established paradigm for studying complex spatio-temporal dynamics and pattern formation in reaction-diffusion processes \cite{markstein1949cell,zeldovich1944theory,moore1995combustion,smith1928bunsen,gololobov1981two,bregeon1978near,quinard1984cellular,hyman1986kuramoto,sivashinsky1979self,brailovsky1993chaotic,frankel1994sequence,munir1989self}.
Experiments with premixed flames show that combustion fronts can exhibit cellular or oscillatory patterns such as traveling waves, pulsations, and spinning fronts \cite{smith1928bunsen,gololobov1981two}. These patterns originate in hydrodynamic effects and thermal/mass diffusional instabilities \cite{zeldovich1944theory,sivashinsky1977diffusional,sivashinsky1983instabilities}. The former mechanism is based, as shown by Darrieus and Landau, on assuming that the mixture density changes due to gas expansion \cite{darrieus1938unpublished,landau1944theory}, whereas the latter mechanism accounts for the competition between transport phenomena. The thermal-diffusional instability mechanism is thus dominant in combustion in systems of condensed solids where hydrodynamic effects are negligible. 

Numerical simulations and experiments on thermo-diffusive instabilities in combustion have reported a wide range of interface instabilities, ranging from cellular fronts \cite{denet1989local} to  turbulent fluctuations far in the non-linear regime \cite{bregeon1978near,quinard1984cellular,hyman1986kuramoto}. In the latter context, Sivashinsky discusses self-turbulizing cells using a non-linear differential equation to model the surface of a flame front in a premixed gaseous fuel \cite{sivashinsky1979self}. Oscillatory and spin modes due to thermal instabilities have also been reported by Merzhanov et al. \cite{merzhanov1981shs}. Other studies have observed a transition from a uniform solution to chaotic pulsations via periodic doubling \cite{brailovsky1993chaotic,frankel1994sequence}. A more complete review of the different types of instabilities reported in these reactive systems can be found in previous studies \cite{munir1989self}. 

\begin{figure}[htbp!]
\centering
\includegraphics[width=0.5\textwidth]{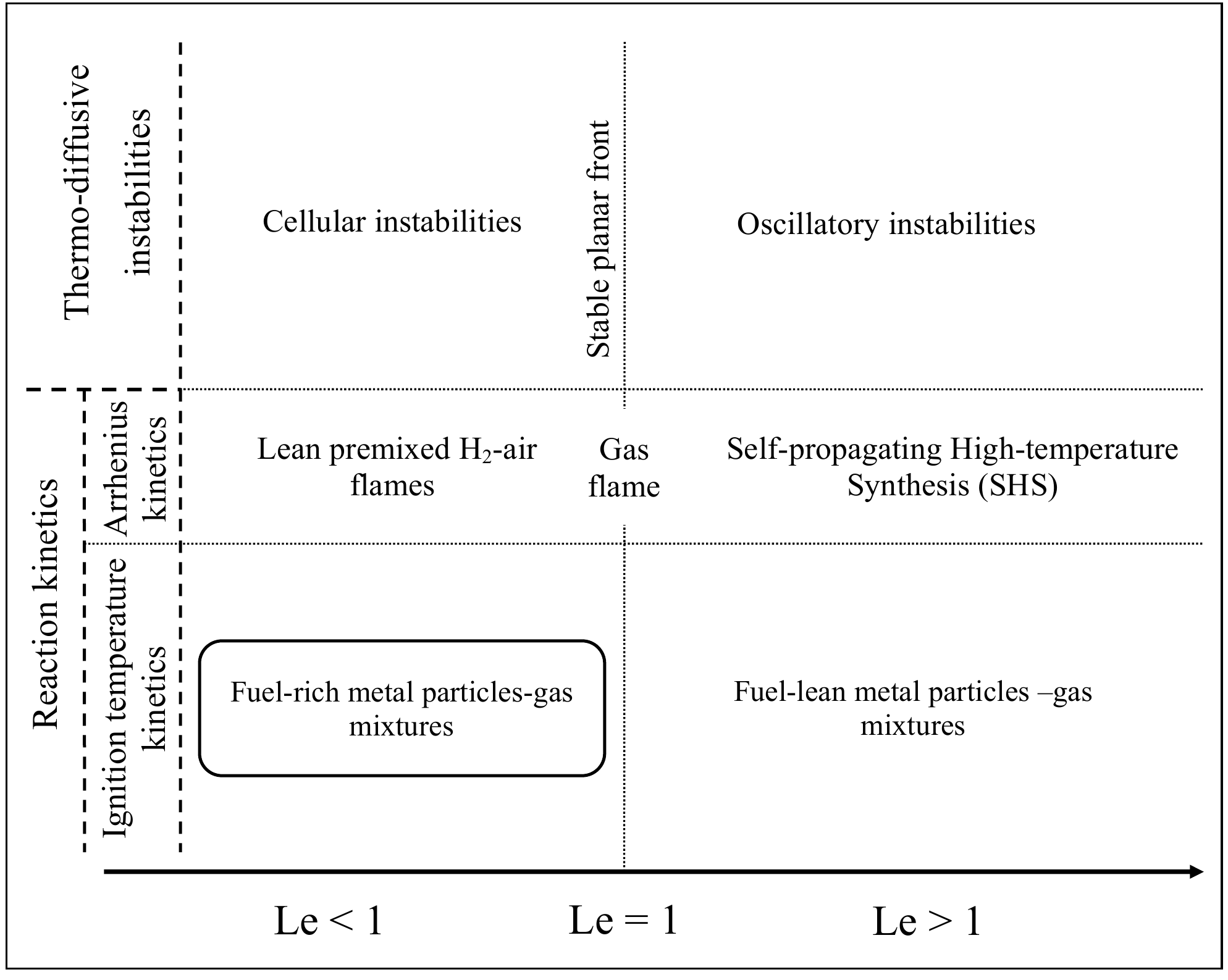}    \\ 
\caption{Classification of different combustion systems with respect to the reaction kinetics and Lewis number. Thermo-diffusive instabilities for these systems in two forms of ``Cells" for low Lewis number values ($\mathrm{Le}<1$), and ``Oscillations" for high Lewis number values ($\mathrm{Le}>1$), have been observed in previous studies \cite{ronney1990near,dimitriou1989dynamics,merzhanov1969theory,julien2015freely}. This paper studies the regime of rich metal particles in gas.} 
\label{fig:parameter-map}
\end{figure}

The mathematics of solid-gas and gas-less solid
\footnote{Gas-less combustion includes a process whereby a solid mixture is converted directly into more stable solid solution, without a gaseous reactant and even without any gaseous product.  
Solid "combustion" of this form is ubiquitous and has many applications in materials science, including in the synthesis of metal alloys, ceramics and super-conductors\cite{hwang1997combustion}}
combustion processes can generally be described in the framework of reaction-diffusion equations that integrate heat and mass diffusion with the detailed chemical and thermodynamic characteristics of the reaction process and reactant materials. In recent years, numerous models have been introduced to describe such combustion processes. While apparently disparate, the various models generally describe such combustion phenomena based on (1) the rate and the order of chemical reaction (oxidation in this case), (2) heat transport and diffusion of oxidizer (and their relative importance, through the Lewis number), and (3) the heterogeneity of the solid medium through which the combustion wave is propagating. Some examples of combustion systems with their appropriate choice of reaction kinetics and observed thermo-diffusive instabilities within a range of Lewis numbers are presented in Fig.~\ref{fig:parameter-map}.

Much of the effort to understand the underlying mechanisms of combustion, and in particular the role of thermo-diffusive instabilities, has focused on systems of ``premixed flames" where the solid reactant distributions are surrounded by gas that reacts with the solid via an Arrhenius-type reaction rate. Even though this choice of kinetics is a qualitatively good description for such systems, Brailovsky et al. \cite{brailovsky2015diffusive} have shown more recently that ignition temperature kinetics is a more realistic choice to describe the reaction. In this mode of combustion, the diffusion of oxidizer towards the solid reactant mixture is rate limiting, effectively ``shutting off" the combustion of the reactant below an {\it ignition}  temperature and ``turning on" the combustion at a {\it constant} rate above the ignition temperature, for a period of time required to fully oxidize the solid reactant. Ignition temperature kinetics have been shown, also recently, to be consistent with the kinetics of oxidation of many metal powders \cite{soo2015reaction}. In this paper we study the thermo-diffusive stability assuming the ignition temperature kinetics, in the low-Lewis number regime.

The first part of this paper begins by introducing a pair of reaction-diffusion equations that conveniently reduce to various previous models for solid-gas combustion. We focus on the special case of the model recently proposed by Brailovsky et al. \cite{brailovsky2015diffusive} due to its quantitative description of the oxidation of solid particulate fuels, which is our particular interest in this work. Specifically, the model assumes ignition temperature kinetics to describe the combustion of solid metal fuel by employing a step-wise activation above an ignition temperature, $T_\mathrm{ig}$. While activation is purely thermal, it also couples to the oxidizer concentration, making the reaction mass-transport-limited \cite{soo2015reaction,goroshin2011reaction,vulis1961thermal}. 

The second part of this paper examines the dynamics of combustion fronts in the parameter space  spanned by the order of reaction ($n$), the Lewis number ($\mathrm{Le}$), and rate of reaction. Specifically, we focus on the role of oxidizer and thermal transport in the development of fronts with cellular and dendritic morphologies in solid-gas combustion, in the limit of continuum fuel sources.  
The linear growth rate and wavelength selection of cellular flame fronts are examined numerically, validating the recent analytical predictions of Brailovsky et al. \cite{brailovsky2015diffusive}. In the non-linear regime, we explore the growth of combustion morphologies that vary from  cells to {\it seaweed} dendrites analogous to those observed in directional solidification  \cite{amoorezaei2010spacing,amoorezaei2012orientation}.  This is seen to happen for low $\mathrm{Le}$, a regime separating cellular/dendritic patterns from oscillatory modes prevalent at high $\mathrm{Le}$. Length scale selection of complex dendritic fingers is analyzed in the context of the linearly unstable $k$ modes available to the system at early times.  
The results of this work are expected to provide a reference for a future study on cellular spacing in systems with randomly distributed fuel particles. 

\section{Solid-Gas Combustion Model}

We model combustion by a simple one-step exothermic reaction assuming constant molecular weight, specific heat capacity $c_{p}$, and a temperature-independent thermal conductivity $\kappa$. Following Sivashinsky, we neglect hydrodynamic effects, and also consider the density of the oxydizer mixture $\rho$ to be uniform in time and space \cite{sivashinsky1983instabilities}. The constant density assumption is plausible based on an argument by Spalding \cite{spalding1956theory} who argued that the density changes (due to gas expansion) are approximately inverse of temperature while the conductivity is changing (approximately) linearly with temperature, which will tend to cancel out the effects of the density change. The reaction rate follows stepwise ignition temperature kinetics, a physically accurate approximation for solid-gas combustion of metals \cite{soo2015reaction,beck2003nonlinear,brailovsky2015diffusive}. We assume that oxidizer transport is diffusive and  ignore the transport of solid fuel as it is much slower that of the oxidizer.
The system is characterized by the temperature field and the oxidizer concentration, and described by the following coupled reaction-diffusion equations,

\begin{flalign}
\left\{
\begin{array}{ll}
\label{eq:evolution}
\frac{\partial{\theta}}{\partial{t}} &= \nabla^2\theta + W \\   
  & \\                     
\frac{\partial{\phi}}{\partial{t}} &= \frac{1}{\mathrm{Le}}\nabla^2\phi - W,       
\end{array}
\right.
\end{flalign}

where $\theta(\vec{x},t) = (T(\vec{x},t)-T_{\mathrm{o}})/(T_{\mathrm{ad}}-T_{\mathrm{o}})$ is the reduced temperature field while $T_{\mathrm{o}}$ and $T_{\mathrm{ad}}$ represent the far-field and adiabatic temperature values, respectively. The reduced ignition temperature is $\theta_{\mathrm{ig}} = (T_{\mathrm{ig}}-T_{\mathrm{o}})/(T_{\mathrm{ad}}-T_{\mathrm{o}})$ where $T_{\mathrm{ig}}$ is the prescribed ignition temperature which determines the onset of reaction of the fuel. We define $\phi(\vec{x},t)$ as the dimensionless local oxidizer concentration ($0\leq\phi\leq\phi_{o}$). The Lewis number ${\mathrm{Le}}$ is defined as the ratio of the thermal diffusivity to the molecular diffusivity, i.e. ${\mathrm{Le}}=\alpha / D$. 
Length and time coordinates of the model are scaled by the characteristic length  $(\delta_{\rm{c}} = \alpha/u)$ and characteristic time $(\tau = \alpha /u^{2} )$, where $u$ is the characteristic velocity of the planar flame front (see Appendix A). 

The source term $W$ in \E{eq:evolution} depends on whether we consider a continuum of fuel (continuum limit) or a discrete set of fuel particles (discrete limit). In the continuum limit, energy release occurs uniformly throughout the mixture, while in the discrete limit, ignition only occurs at the location in a fuel particle. The source term $W$ in the continuum limit is given by:

\begin{flalign}
W= 
\left\{
\begin{array}{ll}
\label{eq:w_continuum}
 {A\, \phi \, \mathrm{H}(\theta-\theta_{\mathrm{ig}})}  &  \hspace{25pt} \mathrm{for}  \  n = 1 \\ 
& \\
 {A \, \mathrm{H}(\theta-\theta_{\mathrm{ig}})\, \mathrm{H}(\phi) } & \hspace{25pt}   \mathrm{for}  \  n = 0,
\end{array}
\right.
\end{flalign}

where $n$ denotes the order of the kinetics. 
We express the continuum combustion model in \E{eq:evolution} in a unified way by re-writing the source term in Eq.~(\ref{eq:w_continuum}) as

\begin{align}
W_{n} &=  A{\phi^{(n)}}\big\lbrack{\rm{H}({\theta(\vec{x},t)}-\theta_{\rm{ig}})\rm{H}^{(1-n)}(\phi(\vec{x},t))}\big\rbrack,    \  n = 0,1
\label{eq:MM2}
\end{align}

Equations~(\ref{eq:evolution}) and (\ref{eq:MM2}) are hereafter referred to as the continuum {\it master model}, the parameters of which are denoted by the 
notation $\mathrm{MM}(n;\mathrm{Le};\theta_{\mathrm{ig}})$, which represent the reaction kinetics order $n$, the Lewis number $\mathrm{Le}$ and the ignition temperature $\theta_{\mathrm{ig}}$, respectively.
The master model can also be specialized to the case of discrete point-like reactants by writing $W_n$ as 

\begin{align}
W_n=
\begin{array}{ll}
\label{eq.w_discrete}
 \sum\limits_{i} A\, \mathrm{\delta}(\vec{x}-\vec{x}_{\mathrm{i}}) \phi^{(n)} \mathrm{H}(\theta-\theta_{\mathrm{ig}}) \, \mathrm{H}^{(n-1)}(\phi),    \  n = 0,1
\end{array}
\end{align}

and the delta functions represent point-like sources localized at random lattice positions $\vec{x}_i$, indexed by $i$. The summations are over all particles, each of which satisfies the ignition condition.   In the models defined by Eq.~(\ref{eq:evolution}) with \E{eq:MM2} or \E{eq.w_discrete}, the temperature/mass reaction conditions described by the model proposed by Brailovsky et al. \cite{brailovsky2015diffusive} (i.e, \textrm{ $\theta \ge \theta_{\mathrm{ig}}$  and $\ \phi>0$}), are encapsulated in the corresponding Heaviside functions. 

\begin{figure}[htbp!]
\includegraphics[width=0.5\textwidth]{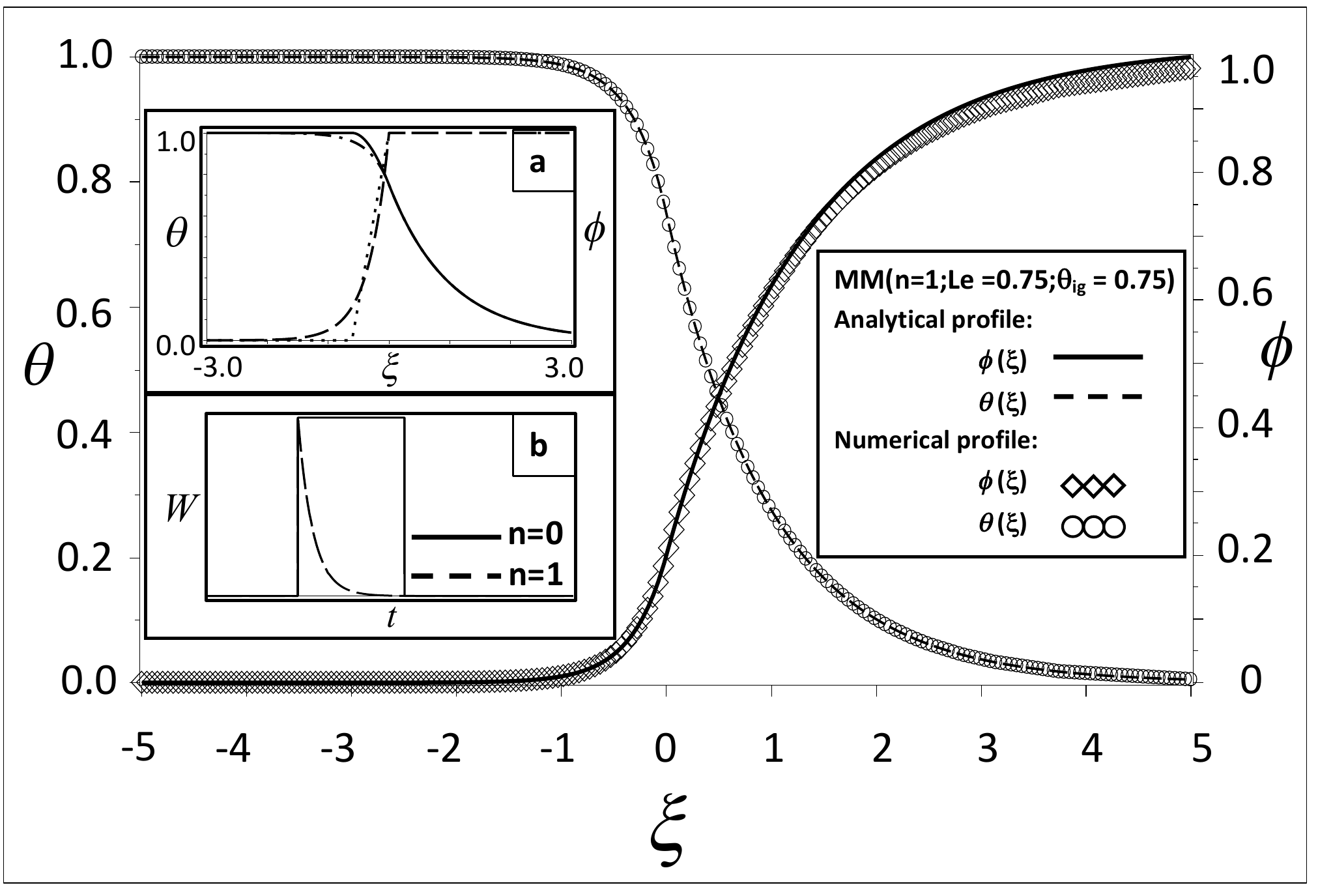}  
\captionof{figure}{	
Numerical and analytical steady-state profiles for $\mathrm{MM}(n=1;\mathrm{Le}=0.75; \theta_{\mathrm{ig}}=0.75)$.
Analytical profiles are hidden under corresponding numerical results.
Inset (a) shows solution profiles for $\rm{MM}$($n=0$; $\rm{Le} = \infty$; $\theta_{\rm{ig}}=0.75$):\,
\textbf{-----} analytical temperature profile; 
\textbf{. . . . } analytical concentration profile;
and for $\rm{MM}$($n=0$; $\rm{Le} =0.75$; $\theta_{\rm{ig}}=0.75$):\, 
\textbf{-.-.-.-.} analytical temperature field;
 \textbf{- - -}   analytical concentration profile.  
Inset (b) shows the schematic source term in time and in the limit of $\rm{Le} \to \infty$ for
$n=0$ and $n=1$ kinetics. 
}
\label{fig:profiles_comp}
\end{figure}

To obtain an estimate of typical parameters for this model, we use experimental data from the combustion of rich-fuel aluminium dust clouds, which give the thermal diffusivity as $\alpha \sim 2\times10^{-5} [\rm{m^2}/\rm{s}]$, and particle reaction time \mbox{$t_{\rm{r}} =\bigO(10^{-3})[\rm{s}]$} \cite{huang2007combustion,beckstead2004summary}. Equation~(\ref{eq:SS-velocity_dimless}) with the assumption of ignition temperature $\theta_{\rm{ig}} =0.75$ and  $\mathrm{Le} =0.75$, yields $u\sim 4\times10^{-2}[\rm{m}/\rm{s}]$ for the characteristic velocity, making the characteristic length $\delta_{\rm{c}}=\alpha/u =\bigO(10^{-4})[\rm{m}]$ and the characteristic time  $\tau =\delta_{\rm{c}}^2/\alpha=\bigO(10^{-2})[\rm{s}]$ for the model's dimensionless time scale. Figure~\ref{fig:profiles_comp} shows typical numerical steady-state profiles of temperature and concentration obtained from the master model with zero and first order kinetics, i.e., $n=0$ and $n=1$, and using the above parameters.

\subsection{Special Limits of the Model}

In this section we analyze the master model, Equations~(\ref{eq:evolution}) and (\ref{eq:MM2}), in the limit of $\rm{Le} \to \infty$ for zero and first order kinetics ($n=0,1$). Both of these cases were previously examined by \cite{sivashinsky1981spinning,matkowsky1978propagation}.
We also discuss briefly a simplification of the master model in the case of discrete reactant sources, when the source term is described by \E{eq.w_discrete}.

\subsubsection{Continuum limit, $\rm{Le} \to \infty$, $n=0$}

In the limit $\rm{Le} \to \infty$ with zero order kinetics the oxidizer equation of the master model can be written as:
\begin{align}
\label{eq.oxevol_cont_0k}
\frac{\partial\phi}{\partial t} &= -A\,\mathrm{H}\big(\theta(\vec{x},t)-\theta_{\rm{ig}}\big)\mathrm{H}\big(\phi(\vec{x},t)\big)
\end{align}
Integrating both sides of \E{eq.oxevol_cont_0k} with respect to the dimensionless time, from $0$ to $\tau_{\rm{r}}$ yields
\begin{align}
\label{eq.Adt}
A=\frac{\phi_{o}}{\bigg\lbrack{\int_0^{\tau_{\rm{r}}}{ \mathrm{H}\big(\theta(\vec{x},t^{'}\big)-\theta_{\mathrm{ig}})\,\rm{H}\big(\phi(\vec{x},t^{'})\big)\,dt^{'}}}\bigg\rbrack},
\end{align}
where $\phi_{o}=\phi(\vec{x},t=0)$ is the initial value of the oxidizer concentration and $\tau_{\rm{r}}=t_{\rm{r}}/\tau$ is the dimensionless particle reaction time. During the particle reaction time, that is, during the time that combustion is ongoing,  \textrm{$\theta \ge \theta_{\mathrm{ig}}$ and $\ \phi>0$}. Equation~(\ref{eq.Adt}) thus becomes $A={\phi_o}/{\tau_{\rm{r}}}$, and the source term on the right hand side of \E{eq.oxevol_cont_0k}, can be written as
\begin{align}
W(\vec{x},t) &=  \frac{\phi_{o}\,\mathrm{H}\big(\theta(\vec{x},t)-\theta_{\mathrm{ig}}\big)\,\mathrm{H}\big(\phi(\vec{x},t)\big)}{{\tau_{\rm{r}}}}.
\label{W2}
\end{align}

Alternatively, by introducing the ignition time $t_{\mathrm{ig}}(\vec{x})$, the time at which the local temperature at position $\vec{x}$ rises above the ignition temperature, one can make the Heaviside functions in \E{W2} depend explicitly on time \cite{beck2003nonlinear}. In that case, the source term takes the form
\begin{align}
\label{eq.w_t_cont_0k}
W(\vec{x},t) &= \frac{\phi_{o}\,\mathrm{H}\big(t-t_{\mathrm{ig}}(\vec{x})\big)\,\mathrm{H}\big(\tau_{\rm{r}}-(t-t_{\mathrm{ig}}(\vec{x}))\big)}{\tau_{\rm{r}}},
\end{align}
and the master model  reduces to a single heat diffusion equation
\begin{align}
\label{eq.evol_t_cont_0k}
\frac{\partial\theta}{\partial t} &= \nabla^2\theta +  \frac{\phi_{o}\,\mathrm{H}\big(t-t_{\mathrm{ig}}(\vec{x})\big)\,\mathrm{H}\big(\tau_{\rm{r}}-(t-t_{\mathrm{ig}}(\vec{x}))\big)}{\tau_{\rm{r}}}.
\end{align}
Equation~(\ref{eq.evol_t_cont_0k}) describes continuum gas-less combustion with ignition temperature kinetics, and is hereafter referred to as the {\it continuum box model}. The term ``box'' derives from it being a continuity equation, i.e, describing transport (of heat) in-and-out of a finite region (a ``box'') located at position $\vec{x}$.

\begin{figure}[!h]
\includegraphics[scale = 0.36]{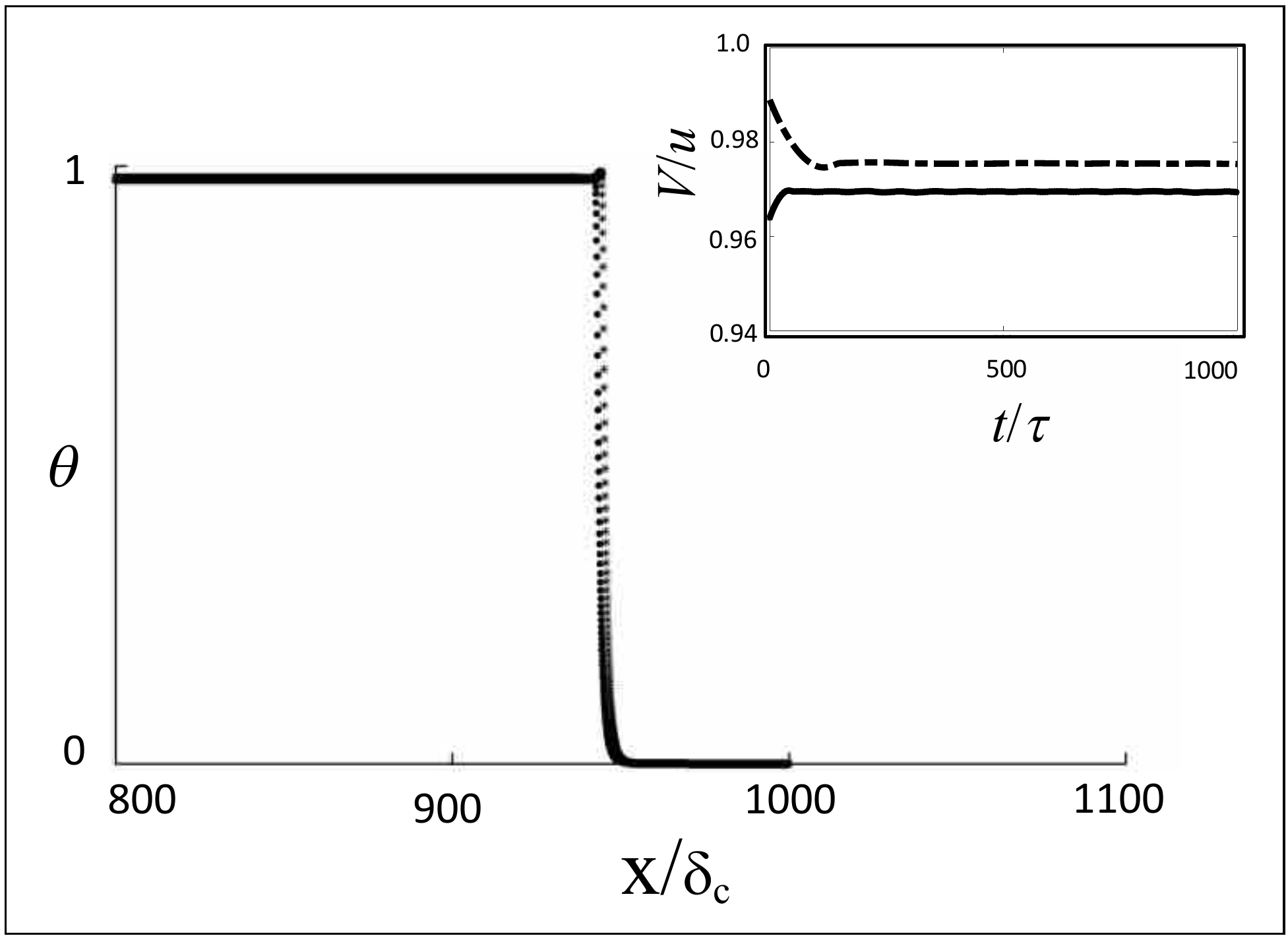}}{\caption{Temperature profile for a zero-order kinetics combustion front in the absence of mass diffusion, simulated with: $(\protect\marksymbol{*}{black})$ master model at $t=967.68\tau$, $(\ast)$ ``continuum box model" at $t=962.56\tau$. The temperature profiles are shifted in time to compare the profiles. Only parts of the profiles around the interfaces are plotted. Inset: average velocity $V$ of the planar front as a function of the simulation time : \---- master model , \- - - - continuum box model.}
\label{fig:MM-velocity}
\end{figure}

The continuum box model corresponds to the limit of the continuum master model with zero-order kinetics in the absence of mass diffusion. Figure \ref{fig:MM-velocity} shows the numerical one-dimensional temperature profile along the principle direction of front propagation ($x$) obtained using the continuum master model \mbox{$\mathrm{MM}(n=0;\mathrm{Le \to \infty};\theta_{\mathrm{ig}}=0.75)$} and using the continuum box model with $\theta_{\rm{ig}}=0.75$ and $\tau_{\rm{r}}=\phi_o/18$ (see Appendix~A).
The inset in Fig.~\ref{fig:MM-velocity} shows the velocity evolution of the combustion front in both cases. A difference in the initial front velocities leads to slightly different transients and steady-state front speeds. This can be avoided by directly using the steady-state solution as the initial condition. 

\subsubsection{Continuum limit, $\rm{Le} \to \infty$, $n=1$}

When considering first order kinetics ($n=1$) and no mass diffusion ($\rm{Le} \to \infty$), the evolution of the oxidizer concentration in equation~(\ref{eq:evolution}) takes the form
\begin{align}
\phi(x,t) &= \phi_{o}\,\mathrm{exp}{\big\lbrack{-\int_0^t A \, \mathrm{H}\big(\theta(\vec{x},t^{'})-\theta_{\rm{ig}}\big)dt^{'}}\rbrack},
\end{align}
The integrand is constant for $0 \le \tau \le \tau_r$, hence $A ={1}/{\tau_r}$ defines the time until the oxidizer reaches $\rm{e}^{-1}$ of its initial value.  Now the master model,  \E{eq:evolution} and (\ref{eq:MM2}), can be reduced to a single heat diffusion equation with a source given by

\begin{align}
W &=  \frac{\phi_{o} \, \mathrm{H}\big(\theta(\vec{x},t)-\theta_{\mathrm{ig}}\big)}{\tau_r}\bigg\lbrack{\mathrm{\exp}^{\big\lbrack{-\int_0^t \frac{\mathrm{H}(\theta\big(\vec{x},t^{'})-\theta_{\mathrm{ig}}\big)}{\tau_r}dt^{'}}\rbrack}}\bigg\rbrack
\label{explicitW_n1}
\end{align}
The qualitative difference between the source in this case (continuum, $\rm{Le} \to \infty$, $n=1$, Eq.~(\ref{explicitW_n1})), and in the previous case (continuum, $\rm{Le} \to \infty$, $n=0$, Eq.~(\ref{eq.w_t_cont_0k})) is illustrated in inset b of Fig.~\ref{fig:profiles_comp}. Both have a finite duration, but while active one is constant and the other decays exponentially.

\subsubsection{Discrete limit of the Box Model}

Considering discrete particle sources described by Eq.~(\ref{eq.w_discrete}) in the limit of zero-order kinetics ($n=0$) and $\rm{Le} \to \infty$, transforms the continuum box model of \E{eq.evol_t_cont_0k} into the discrete box model.
Here, the inter-particle distance $l$ is of the order of the flame front width of a steady-state planar combustion front, and sets the length-scale $\delta_{\rm{c}}$ in our dimensionless equations, while the dimensionless reaction time becomes $\tau_{\rm{r}}={\alpha t_{\rm{r}}}/{l^{2}}$.
A detailed analysis of this model has been done by Tang et al. \cite{tang2009effect}. We will study this regime via the master model in detail in an upcoming paper that will focus on the propagation of a combustion front in a medium of randomly distributed discrete fuel particles.

\section{Numerical Methods}

We performed our numerical simulations using a C++ finite difference adaptive mesh refinement (AMR) code that incorporates OpenMP parallelization. The adaptive mesh is based on an algorithm originally introduced by Provatas et al. \cite{provatas1998efficient,athreya2007adaptive}, which is specially suited for modelling free-boundary problems. It dramatically decreases CPU times and memory usage in such problems when compared to uniform or fixed grid techniques as it scales both metrics with the size of the interface rather than with the dimension of the system. 
The local refinement of the computational mesh is determined by an error estimator that places a threshold on each of the gradients of temperature and concentration. The lowest refinement level, i.e, the smallest grid size, was set to $dx = 0.04$ (in units of the model's characteristic length-scale $\delta_{\rm{c}}$).
The time step was determined according to the numerical stability criteria for the two dimensional diffusion equation, evaluated using the larger gradient coefficient. Parallelization was performed within 16 core nodes. 
The simulation domain consisted of a two dimensional rectangle ($L_{x} , L_{y}$) where the spatial coordinates $x$ and $y$ were parallel and normal to the principal direction of flame propagation, respectively. The AMR combined with parallelization allowed us to attain large domain sizes, minimizing finite size effects. 

We implemented far-field boundary conditions in the principal direction of flame propagation:
\begin{align}
\theta(+\infty , y) =  0,\hspace{25pt} \theta(-\infty , y) = 1, \\ 
\phi(+\infty , y) =  1, \hspace{25pt}  \phi(-\infty , y)  = 0, \nonumber
\label{eq:far_field}
\end{align}
and zero flux boundary conditions in the lateral direction in order to model adiabatic conditions:
\begin{align}
\frac{\partial{\theta}}{\partial{y}}(x, y = 0) =  0,\hspace{25pt} \frac{\partial{\theta}}{\partial{y}}(x, y = L_y) =  0,   \\
\frac{\partial{\phi}}{\partial{y}}(x, y = 0) =  0,\hspace{25pt} \frac{\partial{\phi}}{\partial{y}}(x, y = L_y) =  0. \nonumber
\end{align}
	
\section{Analysis of Cellular Flame Front Instability}

Below a critical value of the Lewis number $\mathrm{Le}_{\rm{c}}\sim \mathcal{O}(1)$, when mass diffusion becomes large enough relative to heat diffusion that the combustion front develops cellular flames. These structures, which have troughs pointing in the direction of the consumed material and convex peaks that grow towards the fresh mixture, span the combustion front in periodic or periodic-like formations \cite{markstein1949cell}. 
In this section we address the linear stability of cellular fronts numerically and compare our results to the recent analytical predictions of Brailovsky et al. \cite{brailovsky2015diffusive}. The intermediate to long-time dynamics and the length-scale selection of cellular flame fronts in the vicinity of $\mathrm{Le}_{\rm{c}}$ are then examined numerically. 

\subsection {Linear regime} 

A linear dispersion relation of a moving front pairs the linear growth rate $\omega$ of a periodic perturbation of a planar interface with the wavenumber $k$ of the perturbation.
Brailovsky et al.  \cite{brailovsky2015diffusive} derived analytically the following dispersion relation for the first order ($n=1$) kinetics of the master model,
\begin{flalign}   
\label{eq:Forder}  
\Delta (\omega,k,\mathrm{Le},\theta_{\rm{ig}}) &\equiv (q-l)\lbrack\theta_{\rm{ig}}-2(1-\theta_{\rm{ig}})p \rbrack && \\\nonumber 
         & \lbrack(1-\mathrm{Le})(\omega-l)- A\,\mathrm{Le} \rbrack +A\,\mathrm{Le}\,(p-l) = 0,&&       
\end{flalign}
while for zero order ($n=0$) kinetics they obtained
\begin{flalign}
\Delta (\omega,k,\mathrm{Le},\theta_{\rm{ig}}) &\equiv 1+(1+2p)\lbrack \exp(-R)-1\rbrack &&\\\nonumber
           & -\mathrm{\exp}\lbrack R(\mathrm{Le}-1+q-p)\rbrack =0,               
 \label{eq:disper-Zorder}           
\end{flalign}

\noindent where $p$, $q$, $l$ are defined by
\begin{flalign}
 p &= \frac{1}{2}(\sqrt{1+4\omega+4k^{2}}-1), &\\
 q &= -\frac{1}{2}(\sqrt{{\mathrm{Le}}^{2}+4\mathrm{Le}\,\omega+4k^{2}}+\mathrm{Le}) ,&\\ 
 l &= \frac{1}{2}(\sqrt{{\mathrm{Le}}^{2}+4\mathrm{Le}(A+\omega)+4k^{2}}-\mathrm{Le}). &
\end{flalign}

Setting $k=0$ in the dispersion relation for  $n=1$ kinetics (Eq.~\ref{eq:Forder}) and solving for $\omega=0$, leads to the critical value of the Lewis number at which $k=0$ becomes neutrally stable \cite{brailovsky2015diffusive},
\begin{flalign}
& \mathrm{Le_{\rm{c}}}(\theta_{\mathrm{\rm{ig}}},k=0) = \nonumber \\
& \frac{\sqrt{16\theta_{\rm{ig}}-15{\theta_{\rm{ig}}}^{2}+2{\theta_{\rm{ig}}}^3+{\theta_{\rm{ig}}}^4}+{\theta_{\rm{ig}}}^2-3\theta_{\rm{ig}}}{(4-6\theta_{\rm{ig}}+2{\theta_{\rm{ig}})^{2}}}.
\end{flalign}  
This value of the Lewis number specifies the boundary between the stable (planar) and cellular regimes as a function of ignition temperature. We hereafter identify this value with, and refer to it as, the critical Lewis number \mbox{$\rm{Le}_{o}$= $\mathrm{Le_{\rm{c}}}(\theta_{\mathrm{\rm{ig}}},k=0).$} 

\begin{figure}[htbp!]
\centering
\includegraphics[width=0.5\textwidth]{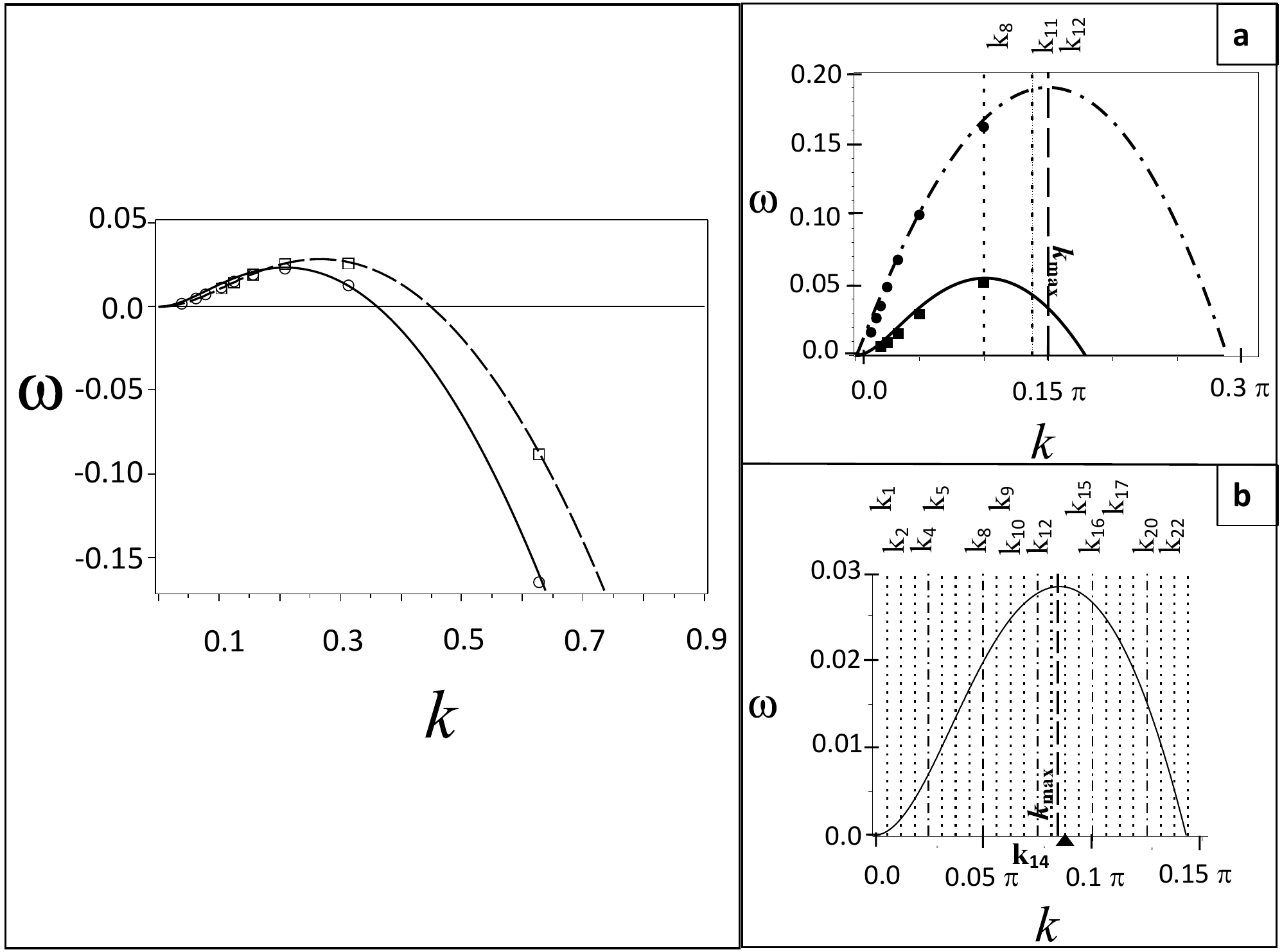} 
\captionof{figure}{Left: Dispersion relation for  $\mathrm{MM}(n=1;\mathrm{Le} = 0.6; \theta_{\mathrm{ig}} =0.75)$: 		
		\textcolor{black}{$ \protect\dashedrule$ analytical,}
              	\textcolor{black}{$\medsquare$ numerical; $\mathrm{MM}(n=0;\mathrm{Le} =0.3;\theta_{\mathrm{ig}}=0.75)$:}           
              	\textcolor{black}{$\solidrule$ analytical,} \textcolor{black}{$\medcircle$ numerical. }
		\\
		Right:
		(a) Dispersion relation for  $\mathrm{MM}(n=1;\mathrm{Le} = 0.5; \theta_{\mathrm{ig}} =0.75)$: 	
		\textcolor{black}{$\protect\solidrule$ analytical,} \textcolor{ black}{\protect\marksymbol{square*}{black} numerical$;
		\mathrm{MM}(n=1;\mathrm{Le} = 0.05; \theta_{\mathrm{ig}} =0.75)$:} \textcolor{black}{$\protect\dashedrule$ 
		analytical,} \textcolor{black}{ \protect\marksymbol{*}{black} numerical. Some permissible $k$ modes for $(n=1;\mathrm{Le} = 0.05; \theta_{\mathrm{ig}} =0.75)$ in a system with	
		\textcolor{black}{ $L_{y} =160\, \delta_{\rm{c}}$: $\cdots$ }   
		(b) Dispersion relation for $(n=1;\mathrm{Le} = 0.6; \theta_{\mathrm{ig}} =0.75)$. Permissible $k$ modes  in a system  with	
		\textcolor{black}{ $L_{y} =80\, \delta_{\rm{c}}$: \-.-.-.- }\textcolor{black}{and $L_{y} =320\, \delta_{\rm{c}}$ : $\protect\dots$} $k_{\rm{max}}$ is specified in both (a) and (b) by dashed line. 
		} }
   \label{fig:dispersion}
\end{figure}

To obtain a linear dispersion relation numerically, we perturb a steady-state planar front solution with a small amplitude sinusoidal wave and compute the growth rate of the perturbation at its initial stages, while the evolution of the perturbation's amplitude is purely exponential.
More explicitly, given a steady-state planar front solution for the dimensionless temperature field $\theta_{\rm{ss}} (\bar{r},t)$ and the oxidizer concentration $\phi_{\rm{ss}} (x,t)$, we use as initial condition $\theta_{\rm{ss}}(x-A(y),t)$ and $\phi_{\rm{ss}}(x-A(y),t)$ with $A(y)=A_{\rm{o}}\cos (k y)$ where $A_{\rm{o}}$ is the amplitude of the perturbation, $k$ is its normal wavenumber and $\lambda=2\pi /k$ its wavelength. The growth rate is defined by $\omega=(1/A) dA/dt $, and the regime is considered linear as long as the growth rate increases linearly in time.
Linear dispersion calculations were performed for different transverse wavenumbers for both zero order kinetics $\mathrm{MM}(n=0;\mathrm{Le} = 0.3;\theta_{\rm{ig}} =0.75)$ and first order kinetics $\mathrm{MM}(n=1;\mathrm{Le} = 0.6;\theta_{\rm{ig}} =0.75)$. The lateral dimension of the system was set to half of the perturbation's wavelength, and was varied according to the wavelength analyzed. The resulting dispersion computations are shown in Fig.~\ref{fig:dispersion} (main). The respective analytical solutions of Brailovsky et al., shown in the same figure, are in excellent agreement.

In Fig.~\ref{fig:dispersion} (main), the dispersion relation for first order kinetics \mbox{($n=1$)} has a larger range of unstable $k$ modes than that for zero order ($n=0$) kinetics, as well as a larger maximum growth rate. This is because with zero order kinetics ($n=0$), the rate of heat release throughout the reaction zone is constant, making the front more stable against perturbations than in the corresponding first order kinetics \mbox{($n=1$)} case.
The same effect, i.e, increasing the range of unstable $k$ modes and the maximum growth rate, can be achieved by decreasing the Lewis number while holding all other parameters fixed, as illustrated in Fig.~\ref{fig:dispersion}(a). 
Therefore, one can make the first order kinetics \mbox{($n=1$)} case comparable to the zero order kinetics ($n=0$) case by increasing the mass transport rate or by decreasing the heat diffusion.

\subsubsection{Effect of system size on linear mode selection}

\begin{figure}[htbp!]
\centering
\includegraphics[width=0.5\textwidth]{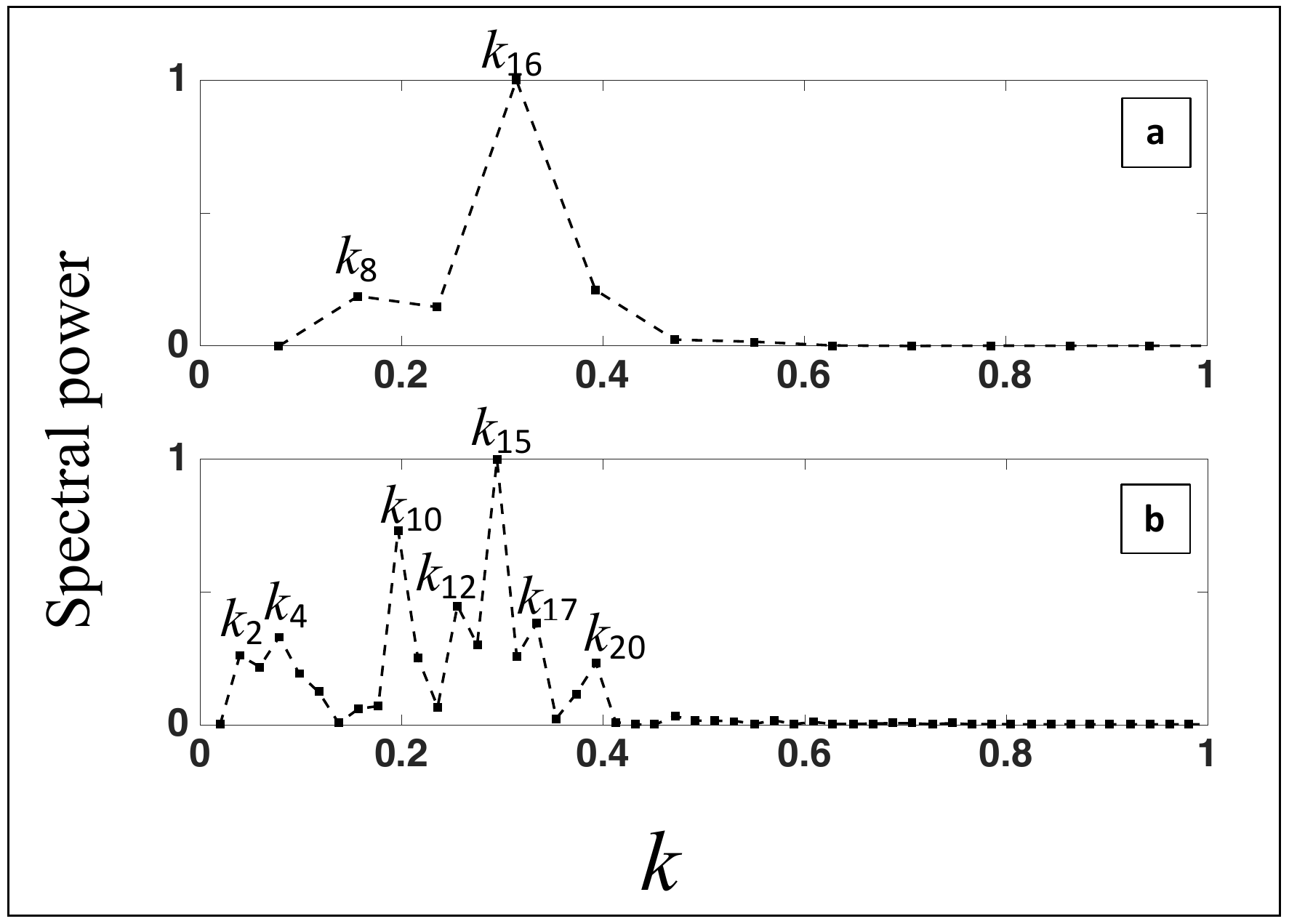}    \\ 
\caption{The normalized spectral power of the 1D interface $x=f(y,t)$ with model parameters $\mathrm{MM}(n=1;\mathrm{Le}=0.6; \theta_{\mathrm{ig}}=0.75)$. The un	stable $k$ modes in the linear regime under noisy initial conditions are shown for lateral domain sizes corresponding to (a) $L_{y} =80\, \delta_{\rm{c}}$ at $ t = 76\tau$ , (b) $L_{y} =320\, \delta_{\rm{c}}$ at $ t = 69\tau$.  }
\label{fig:cell_emergence_noisyints_PSD}
\end{figure}

To examine the effect of system size on growth modes, we consider the evolution of an initially noisy front by perturbing a planar steady-state solution with local point-to-point random morphological fluctuations whose amplitude varies between $0$ and $A_{\rm{o}}$. It is noted that given the non-flux boundary conditions in the transverse $y$-direction of the system, only modes that are in multiples of the width of the system, $L_y$, $2 L_y$, $4L_y$, ... are attainable. 
Some of the accessible modes available for two systems of width $L_{y}=80\, \delta_{\rm{c}}$  and $L_{y}=320 \, \delta_{\rm{c}}$ $(n=1;\mathrm{Le} = 0.6;\theta_{\mathrm{ig}} =0.75)$ are displayed in the dispersion relation in Fig.~\ref{fig:dispersion}(b).  The actual selected modes at early times for these two systems  sizes  are shown in Fig.~\ref{fig:cell_emergence_noisyints_PSD}.

Figure~\ref{fig:cell_emergence_noisyints_PSD} highlights the correlation between the number of potential unstable $k$ modes and the transverse domain size. Two modes, $k_{8} =0.05\pi/\delta_{\rm{c}}$ and $k_{16} =0.1\pi/\delta_{\rm{c}}$, are excited in the smaller system $L_{y} =80\,\delta_{\rm{c}}$. The highest of these two peaks ($k_{16}$) is close to the most unstable mode  $k_{\rm{max}}$ shown in the dispersion relation in Fig.~\ref{fig:dispersion}(b), while the other is 
double the wavelength. Both are, as expected, multiples of the width of the system.
With the wider system,  the number of modes excited is larger and the highest peak ($k_{15}$) is closer to $k_{\rm{max}}$ of Fig.~\ref{fig:dispersion}(b).
The larger the transverse system size $L_y$, the more dense is the set of excitable modes attainable. Thus, the number of modes that can be potentially excited becomes larger, with the fastest growing mode becoming closer to the analytical prediction ($k_{\rm{max}}$ of Fig.~\ref{fig:dispersion}(b) ) in the bigger system.
Our numerical results are also consistent with other studies on the effect of system size on the mode selection mechanism \cite{denet1992numerical,clavin1985dynamic}.

\subsubsection{Effect of Lewis number on linear mode amplitudes}

\begin{figure}[htp!]
\includegraphics[width=0.5\textwidth]{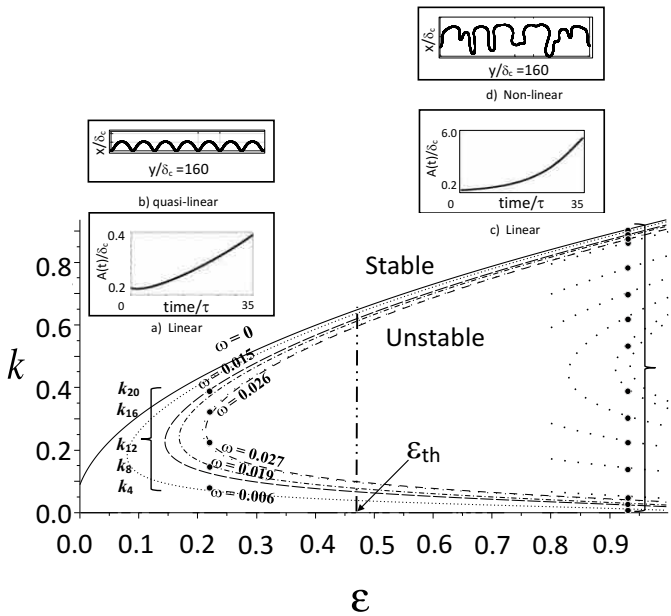}  
\caption{Stability space for $\mathrm{MM}(n=1;\mathrm{Le};\theta_{\mathrm{ig}} =0.75)$ for $\mathrm{Le} < \mathrm{Le}_{o}$ where the parameter $\varepsilon$ is defined as $(\mathrm{Le}_o-\mathrm{Le})/\mathrm{Le}_o$. Curves which are obtained using Eq.~\ref{eq:Forder} define contours of fixed growth rate ($\omega\geq 0$) of perturbations with transverse $k$ mode. Attainable normal modes for system size $L_{y} =80\,\delta_{\rm{c}}$ for $\mathrm{Le} = 0.6 \mathrm\, {\rm and}\, \mathrm{Le} = 0.05$) are denoted by $\protect\marksymbol{*}{black}$. Top-Left inset: (a) and (b) show the evolution of the amplitude and a typical late-time morphology of an interface that was initially perturbed with a sinusoidal wave with $k=0.05\pi/\delta_{\rm{c}}$, for the case of $\mathrm{Le} =0.6$. Top-Right inset: (c) and (d) show the evolution of the amplitude and a typical late-time morphology of a sinusoidally perturbed interface with $k=0.05\pi/\delta_{\rm{c}}$ after a long time in its evolution, for the case $\mathrm{Le} =0.05$. $\rm{\varepsilon}_{\rm{th}}$ in the figure specifies the border between the shallow amplitude (left side) and the large amplitude (right side) cellular regime. }
\label{fig:Linear_weakly_non_linear}
\end{figure}

The stability of the combustion front with respect to the Lewis number is shown in 
Fig.~\ref{fig:Linear_weakly_non_linear} for the case of \mbox{$\mathrm{MM}(n=1;\mathrm{Le};\theta_{\mathrm{ig}} =0.75)$}.
The curves are obtained using Eq.~(\ref{eq:Forder}). Each curve corresponds to a given growth rate, thus the intersection of a $k={\rm const}$ line with a given curve gives the growth rate of that $k$ mode at a specific $\mathrm{Le}$. The range of unstable $k$ modes narrows with the Lewis number value.
As discussed earlier, the number of attainable transverse $k$ modes is limited by the system transverse size (Fig.~\ref{fig:dispersion}(b)). The discrete $k$ modes obtained using a system with a lateral size $L_{y} =80\, \delta_{\rm{c}}$, for two different values of ${\rm Le}$, are marked ($\protect\marksymbol{*}{black}$). 

Figure~\ref{fig:Linear_weakly_non_linear} also includes the evolution of the amplitude and a typical late-time morphology for two different values of the Lewis number (${\rm{Le}=0.05 ,\, \rm{Le}=0.6}$) and the same initial perturbation mode $k=0.05\pi/\delta_{\rm{c}}$.
After an initial transient period, the amplitude in both cases grows linearly with time. It is noteworthy that for the smaller values of ${\rm Le}$, the amplitude of the perturbation grows faster and thus the system will transit more rapidly to what is expected to be a strongly non-linear regime.
Conversely, as ${\rm Le} \rightarrow {\rm Le}_{\rm{o}}$ the growth rate drops. Systems with ${\rm Le}$ in this range take longer to transition from the linear to the non-linear regime, and they are expected to exhibit only weakly non-linear behaviour, even at late times. 

\subsection{Quasi-Linear Regime as $\mathrm{Le} \to \mathrm{Le}_{\rm{o}}$}

The results of the last section suggest that as $\mathrm{Le} \to \mathrm{Le}_{\rm{o}}$ the amplitude governing fluctuations of the combustion front will evolve slowly from zero at intermediate to late times, possibly even saturating at very late times.  As a result, the transition of the combustion front from a linear to non-linear regime will be long, making it plausible that its dynamics, in this parameter regime, can be analyzed over long time periods in terms of the evolution of the linearly unstable $k$ modes. This intermediate regime, which we call here the {\it quasi-linear regime}, is further examined below.

\subsubsection{Effect of initial condition and system size on the dynamics of the interface}

We first consider a planar steady-state front with an initial sinusoidal morphological perturbation, for two systems that have the same length $L_{x} = 2000 \, \delta_{\rm{c}}$ but different widths, given by $ L_{y} =80\, \delta_{\rm{c}}$ and $ L_{y} =320\, \delta_{\rm{c}}$, respectively.
For visualization purposes the figures presented focus on the front and only show a partial view of the whole system that includes a range of $30\, \delta_{\rm{c}}$ around the front.
For the initial sinusoidal perturbation we choose $k$ modes within the corresponding linear unstable range that are commensurate with the width of the system.

\begin{figure}[!h]
\centering
 \includegraphics[scale = 0.45]{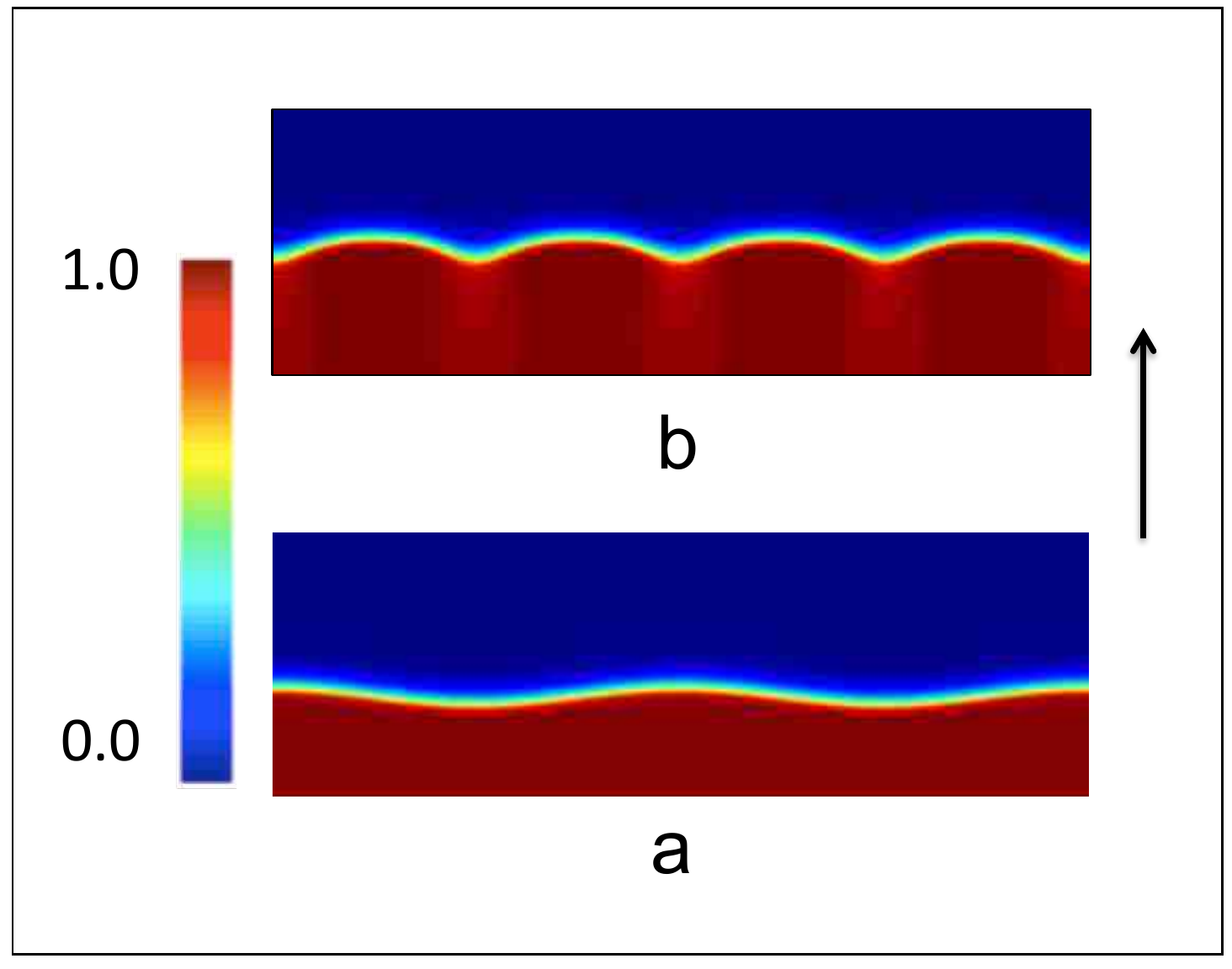}}{\caption{Contours of the dimensionless temperature field in a moving frame for model parameters $\mathrm{MM}(n=1;\mathrm{Le}  = $ 0.6;$\theta_{\mathrm{ig}} = $ 0.75) and system size $L_y=80\,\delta_{\rm{c}}$. a) initial perturbation wavelength $\lambda = $ 40  $\, \delta_{\rm{c}} $ ($k_{8}=0.05\pi/\delta_{\rm{c}}$), b) steady-state front configuration. The arrow shows the direction of  propagation.}\label{fig:Ly80non_lin_cells_lam_forty}
  \end{figure}

Figure \ref{fig:Ly80non_lin_cells_lam_forty} shows two instances in the evolution of the combustion front for model parameters \mbox{$\mathrm{MM}(n=1;\mathrm{Le=0.6};\theta_{\mathrm{ig}}=0.75)$}, a system with $L_y=80\,\delta_{\rm{c}}$, and an initial pure $k$-mode perturbation corresponding to $k_{8} =0.05\pi/\delta_{\rm{c}}$ (see Fig.~\ref{fig:dispersion}(b)). The full simulation run corresponds to $t=1900 \,\tau$, and steady-state is reached after $t =70\,\tau$. 

\begin{figure}[htbp!]
\includegraphics[width=0.28\textwidth]{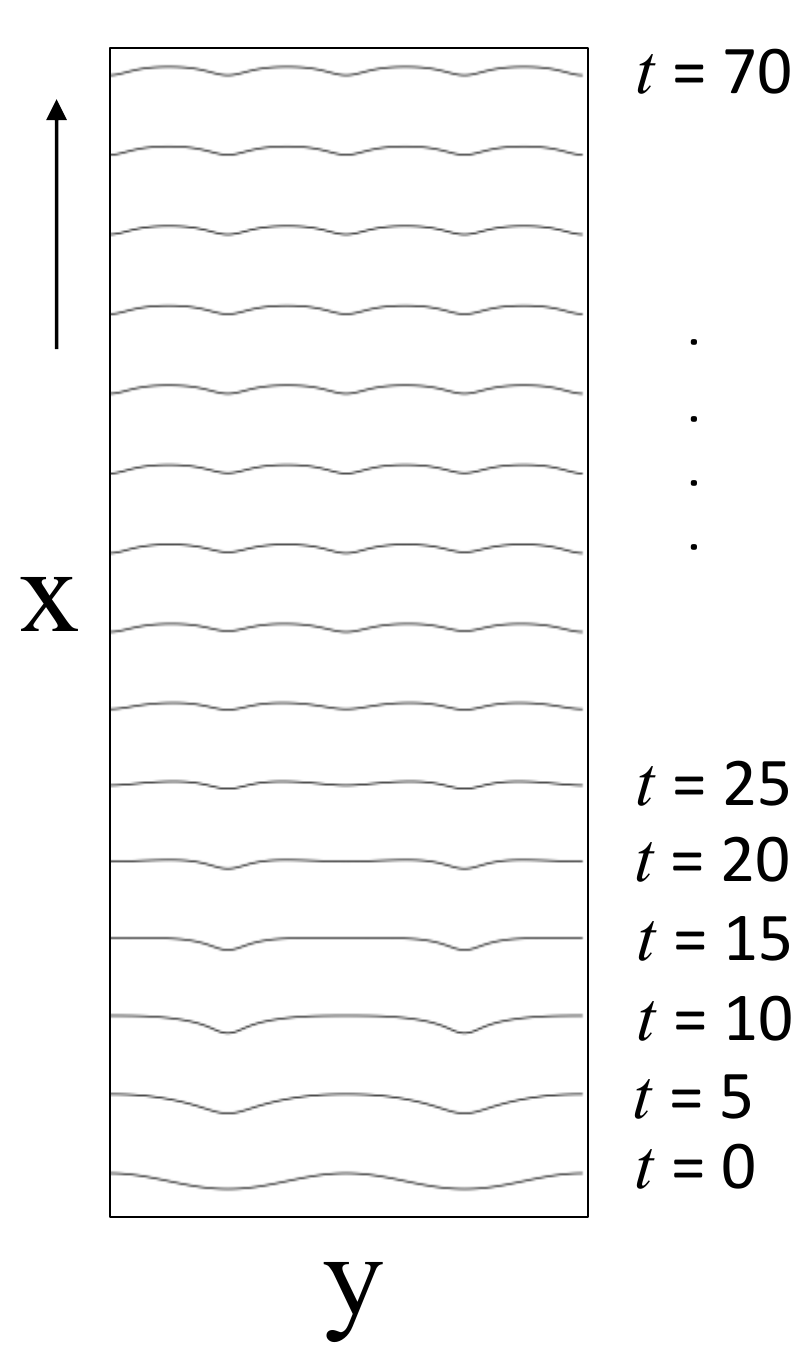}    
\caption{The sequence of the time evolution of the front, defined by the isotherm $\theta(x,y) = \theta_{\mathrm{ig}} = 0.75$ in Fig.~\ref{fig:Ly80non_lin_cells_lam_forty}. Time is in units of time-scale $\tau$ and the isotherm is plotted every $5\tau$. The arrow defines the direction of propagation.}\label{fig:cell_time}
\end{figure}

Figure \ref{fig:cell_time} tracks the evolution of the interface shown in Fig.~\ref{fig:Ly80non_lin_cells_lam_forty}, defined via the temperature isotherm \mbox{$\theta(x,y)=\theta_{\mathrm{ig}}=0.75$}.
The initial sinusoidal perturbation ($k_{8}=0.05\pi/\delta_{\rm{c}}$) flattens due to the high temperature gradient at the tips.
Concurrently, troughs become more pronounced as the cells evolve towards their steady-state shape.
Deep grooves form at approximately $t=5 \tau$, followed by the flattening of the cell tips signalling an upcoming cell splitting event. Around $t=25\,\tau$, the flattened segments become unstable, eventually splitting into two new identical cells and the whole front settles into a steady-state cellular front with half the wavelength of the initial perturbation, $k=k_{16}=0.1\pi/\delta_{\rm{c}}$. This mode is still within the linear unstable range for this system, but much closer than the initial perturbation to the most unstable linear mode $k_{\rm{max}}$ predicted by the dispersion relation in Fig.~\ref{fig:dispersion}(b).

The morphological evolution just described is caused by an interplay between an instability driven by diffusive transport and the effect of surface curvature, as discussed by Zeldovich et al. \cite{zeldovich1944theory}.

Tip splitting as a route for achieving a stable cell array in combustion fronts has been reported in experiments \cite{tabeling1987instability,mitani1980studies} as well as in numerical studies \cite{denet1992numerical,joulin1983dynamics,sivashinsky1979self,sivashinsky1990intrinsic}. Tip splitting as a microstructure selection mechanism is also well known in the context of crystal growth \cite{mullins1963morphological,mullins1964stability, burden1974cellular,burden1974cellular2,langer1980instabilities,trivedi2002interface,greenwood2004crossover,haxhimali2006orientation,gurevich2010phase}. 

\begin{figure}[!h]
\centering
\includegraphics[scale = 0.49]{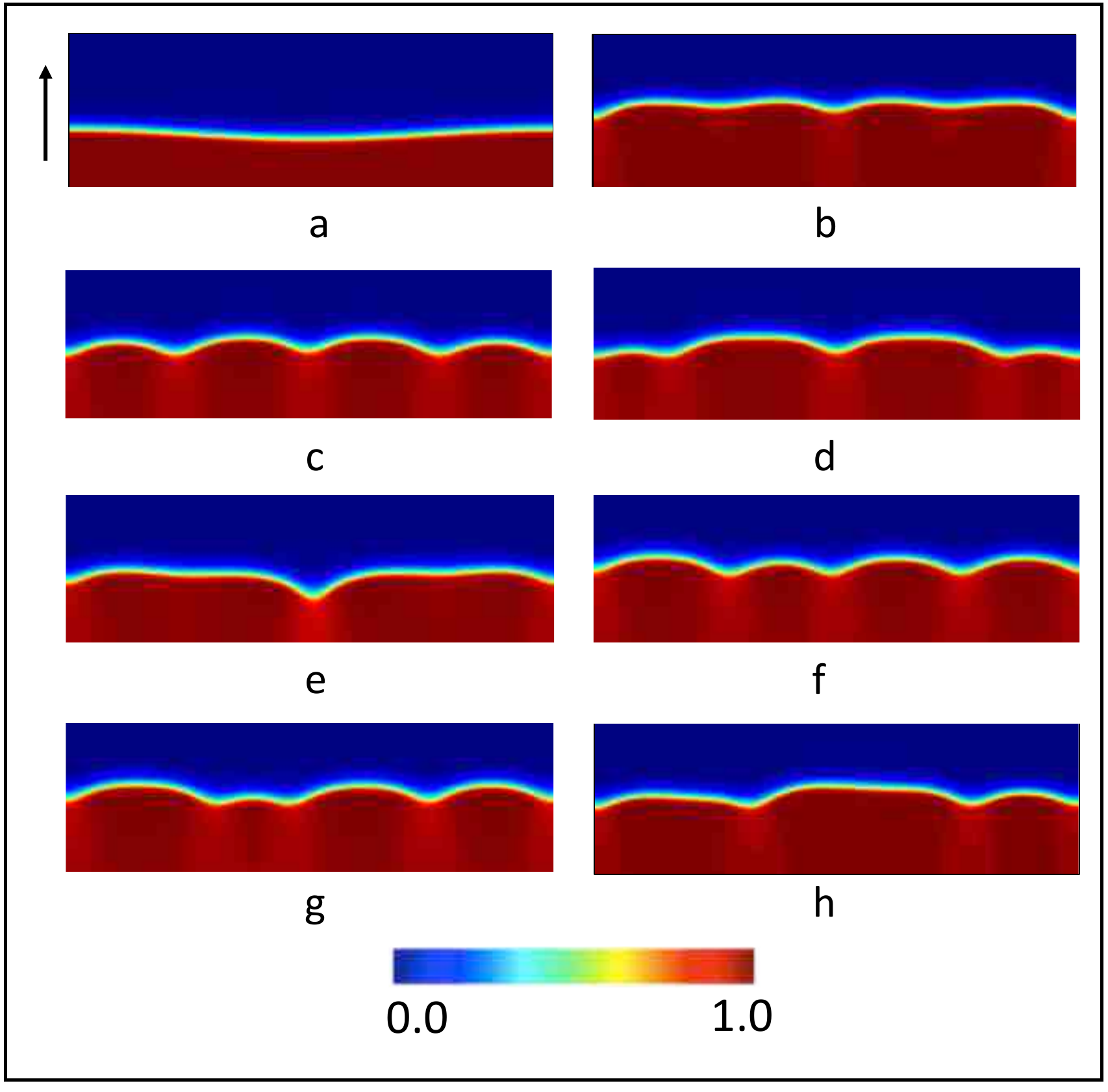}}{\caption{Evolution of the dimensionless temperature front for model parameters $\mathrm{MM}(n=1;\mathrm{Le}  = $ 0.6;$\theta_{\mathrm{ig}} = $ 0.75) from (a): the initially perturbed front with wavelength $\lambda = $ 80 $\, \delta_{\rm{c}} $ to (h): the final state in the computation. The arrow shows the direction of  propagation.}
\label{fig:Ly80non_lin_cells_lam_eighty}
\end{figure}

Figure \ref{fig:Ly80non_lin_cells_lam_eighty} shows the evolution of a flame front with the same model parameters and system size as in Fig.~\ref{fig:Ly80non_lin_cells_lam_forty}, but with an initial pure $k$-mode perturbation corresponding to \mbox{$k_{4} =0.025\pi/\delta_{\rm{c}}$}  (see Fig.~\ref{fig:dispersion}(b)).
Here, we observe that the initial front in (a) undergoes several cell-merging and tip-splitting events and, at least to the end of our simulation in (h) at $t=1900\, \tau$, the system does not reach steady-state. 
Note that the initial perturbation mode $k=0.025\pi/\delta_{\rm{c}}$ is equal to $2\pi/L_y$, the smallest wavenumber allowed by this system size. 

In Fig.~\ref{fig:Ly80non_lin_cells_lam_eighty}(a)  the initial sinusoidal perturbation has a wavelength equal to the width of the system.
Thus, the initial perturbation, which is both very small in the context of the dispersion relation and theoretically the lowest attainable $k$ in the linearly unstable regime for this system size, fails to select a unique wavelength in this range as it is overrun by higher order finite-size effects. Consequently, it does not reach a steady-state. As we can see in Fig.~\ref{fig:Ly80non_lin_cells_lam_eighty}, rather than developing a pure sinusoidal mode compatible with the system size, the front develops a combination of different wavelengths whose competition leads to successive local cell-merging and tip-splitting events but does not settle into a single mode steady-state. It is possible that numerical fluctuations play a role, but in any case, these are smaller than the thermal fluctuation one would expect in a real physical system.
We expect that the same model parameters and initial condition in a wider system can lead to a single mode steady-state corresponding to one of the faster growing allowable modes in the linear regime, after undergoing one or more tip splitting instabilities.

\begin{figure}[!h]
\centering
\includegraphics[width=0.35\textwidth]{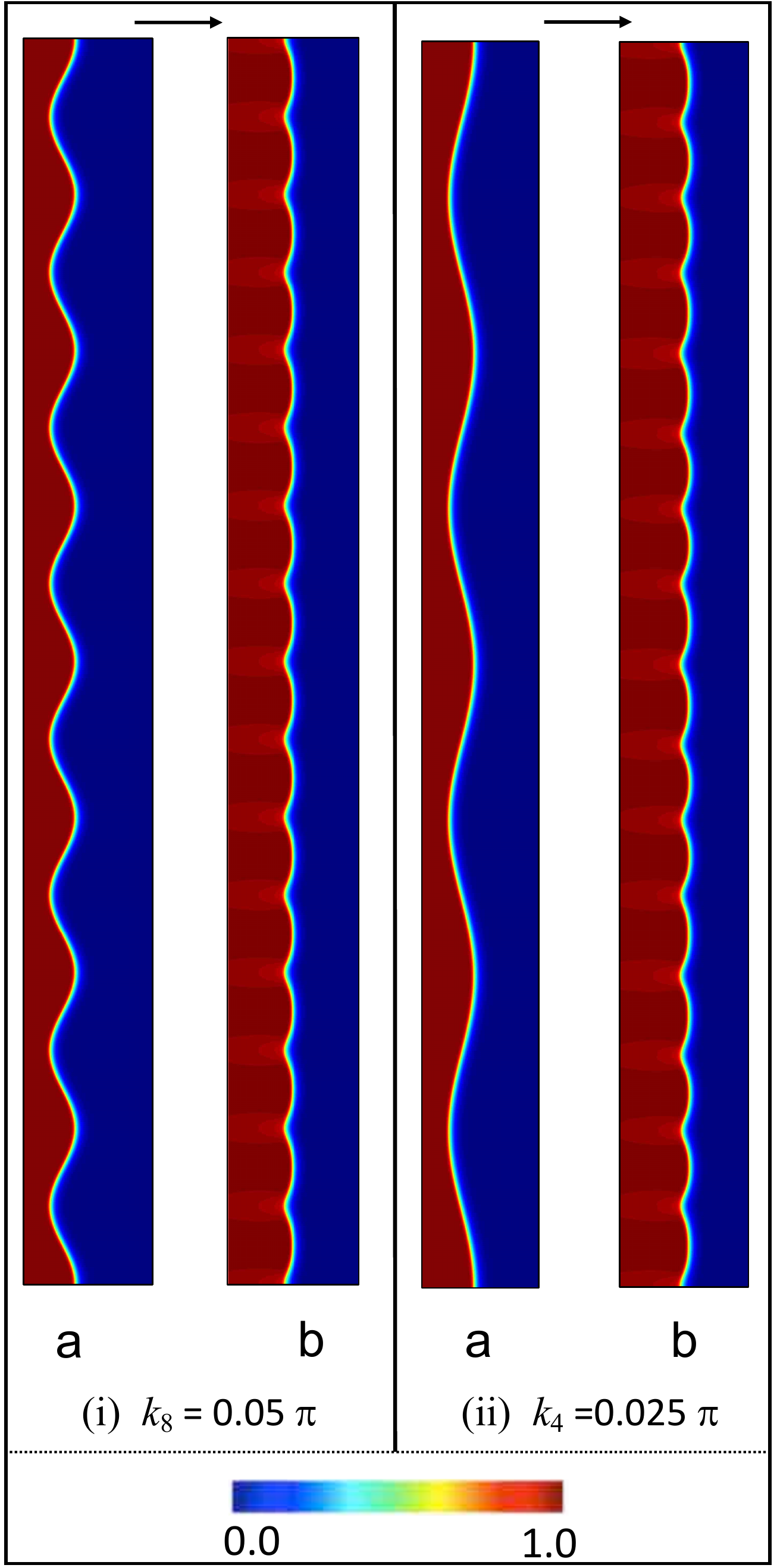}  
\caption{Contour plots of temperature for model parameters $\mathrm{MM}(n=1; \mathrm{Le}=0.6; \theta_{\mathrm{ig}} =0.75)$ in the system $L_{x} =2000$ and $L_{y} =320$. Left: a) initial perturbation with $k =0.05\pi/\delta_{\rm{c}}$  b) steady-state. Right: a) initial perturbation with $k =0.025\pi/\delta_{\rm{c}}$  b) steady-state}
\label{fig:2000_Ly320_lam80_40}
\end{figure}

We next consider the same model parameters as in Fig.~\ref{fig:Ly80non_lin_cells_lam_eighty}, with an initially perturbed interface having $k_{4}$ or $k_{8}$ modes in a wider system of lateral size $L_{y} = 320 \,\delta_{\rm{c}}$. Two instances of the late-time evolution of the front, for each initial condition, are presented in Fig.~(\ref{fig:2000_Ly320_lam80_40}). In both cases the system reaches a steady-state with wavelength $k_{14}$ (in Fig.~\ref{fig:dispersion}(b)), which is close to the steady-state wavelength $k_{16}$ reached by the smaller system (Fig.~\ref{fig:Ly80non_lin_cells_lam_forty}(b)) with an initial perturbation of $k=k_8=0.05\pi/\delta_{\rm{c}}$. As expected, when considering the larger system,  both cases lead to a single mode steady-state corresponding to a fast growing mode in the linear regime of the model, after undergoing several tip splitting events. 

\begin{figure}[!h]
\centering
\includegraphics[width=0.46\textwidth]{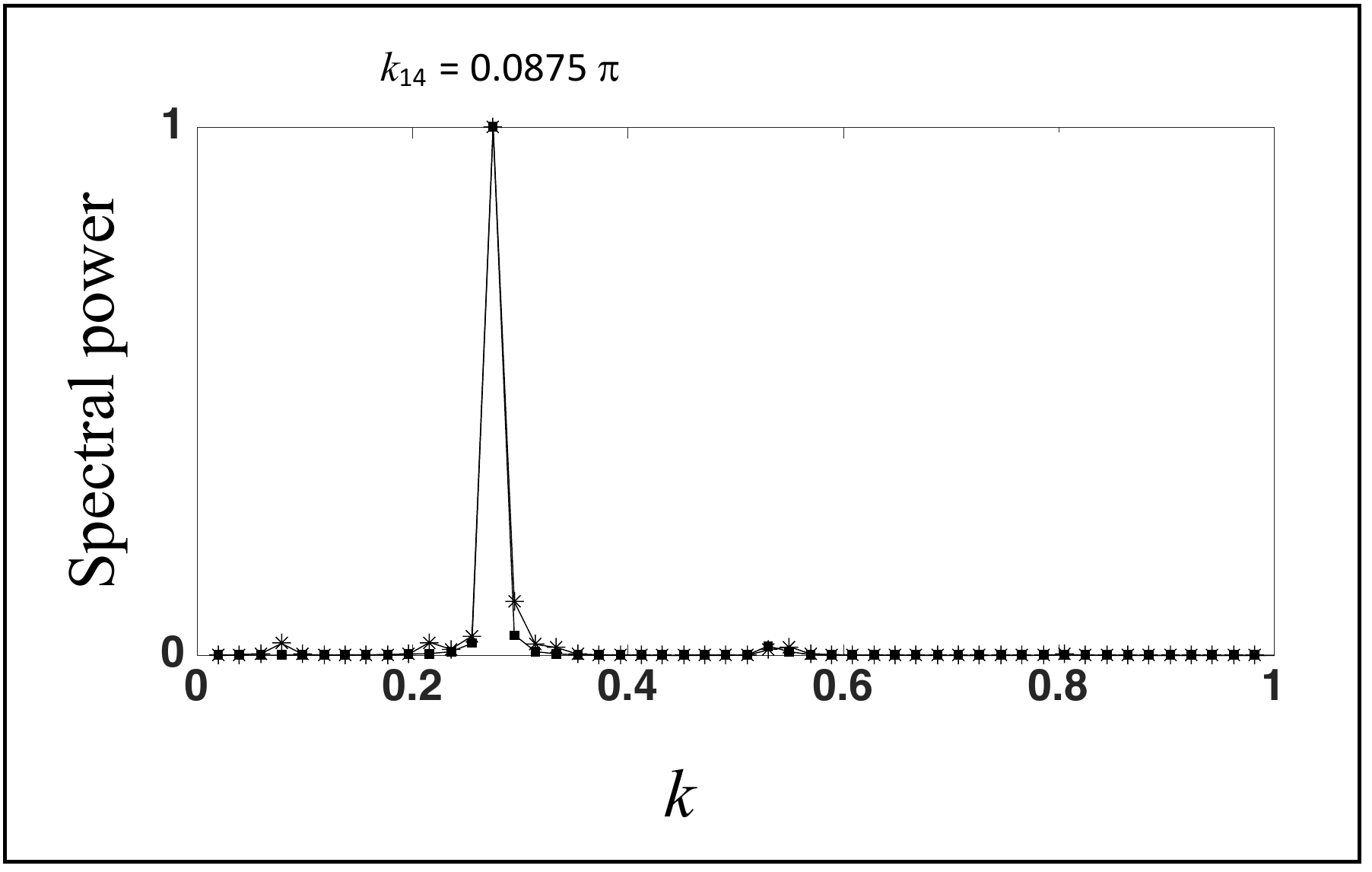}  
\caption{The normalized spectral power of the interface $x=f(y,t)$ of the steady-state for the two different initial perturbation modes in Fig.~\ref{fig:2000_Ly320_lam80_40}: $ k_{8}=0.05 \pi \ (\protect\marksymbol{square*}{black})$ and $k_{4} =0.025\pi/\delta_{\rm{c}}  (\ast)$. }
\label{fig:2000_Ly320_lam80_40_psd}
\end{figure}

The spectral power density of the interface, after reaching the steady-state for the conditions of Fig.~\ref{fig:2000_Ly320_lam80_40}, is shown in Fig.~\ref{fig:2000_Ly320_lam80_40_psd}. The steady-state mode selected, $k_{14}$ (marked by \protect\marksymbol{triangle*}{black} in Fig.~\ref{fig:dispersion}(b)), is remarkably close to the value of  $k_{\rm{max}}$ predicted by the dispersion relation. The predilection of the system to settle into a steady-state with a single mode close to the mode with the highest growth rate predicted by the linear stability analysis was further confirmed by simulating the same model parameters and system size $L_y =320 \,\delta_{\rm{c}}$ with an initial perturbation mode equal to $k_{14}$, which resulted in a steady-state pattern identical to the ones in Fig.~\ref{fig:2000_Ly320_lam80_40} (left/right). Note that the steady-state mode selected in the case of the larger system ($k_{14}$) is closer to the mode with the highest growth rate predicted by the linear stability analysis ($k_{\rm{max}}$) than the steady-state mode selected in the case of the smaller system ($k_{16}$). The higher density of allowable modes in the vicinity of $k_{\rm{max}}$ available to the larger system likely accounts for this wavelength selection mechanism.


It is instructive to examine the evolution of the front for an initial morphologically noisy interface, which contains a combination of many modes and thus can be used to examine their interaction. 
For this purpose, we examine the evolution of an initially noisy front, obtained by perturbing a planar steady-state solution with a local point-to-point random morphological fluctuation with small amplitude for computational domains with $L_{y} = 80\,\delta_{\rm{c}}$ and $L_{y} = 320\,\delta_{\rm{c}}$. 
To study the long-time dynamics of the system, in both cases, a sufficiently large simulation box in the growth direction was chosen, with $L_{x} =5000\,\delta_{\rm{c}}$. 
The duration of the simulation for smaller and larger system size was $4580\,\tau$ and $5500\,\tau$, respectively.

\begin{figure}[!h]
\centering
\includegraphics[scale = 0.5]{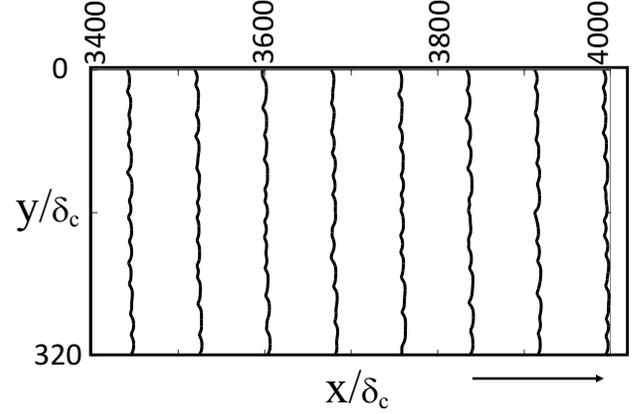}}{\caption{Temporal history of evolution of 1D front corresponding to isotherm $\theta(x,y) =\theta_{\mathrm{ig}}$ for model parameters $\mathrm{MM}(n=1;\mathrm{Le}=0.6;\theta_{\mathrm{ig}}=0.75)$ with initial morphologically noisy interface and in the system size $L_x=5000\,\delta_{\rm{c}} \times L_y=320\,\delta_{\rm{c}}$. The isotherm is plotted every $\sim82\tau$. The arrow shows the direction of  propagation.}\label{fig:5000_320_Le06_noisy}
\end{figure}

Here we find that the front morphology does not converge to steady-state in either case. Instead, the front develops a state where cells are randomly created and annihilated due to localized tip-splitting and cell merging. A partial view of the resulting temporal history of a 1D front $x=f(y,t)$ in the larger domain ($L_y =320\,\delta_{\rm{c}}$) is shown in Fig.~\ref{fig:5000_320_Le06_noisy}. 

\begin{figure}[!h]
\centering
\includegraphics[scale = 0.45]{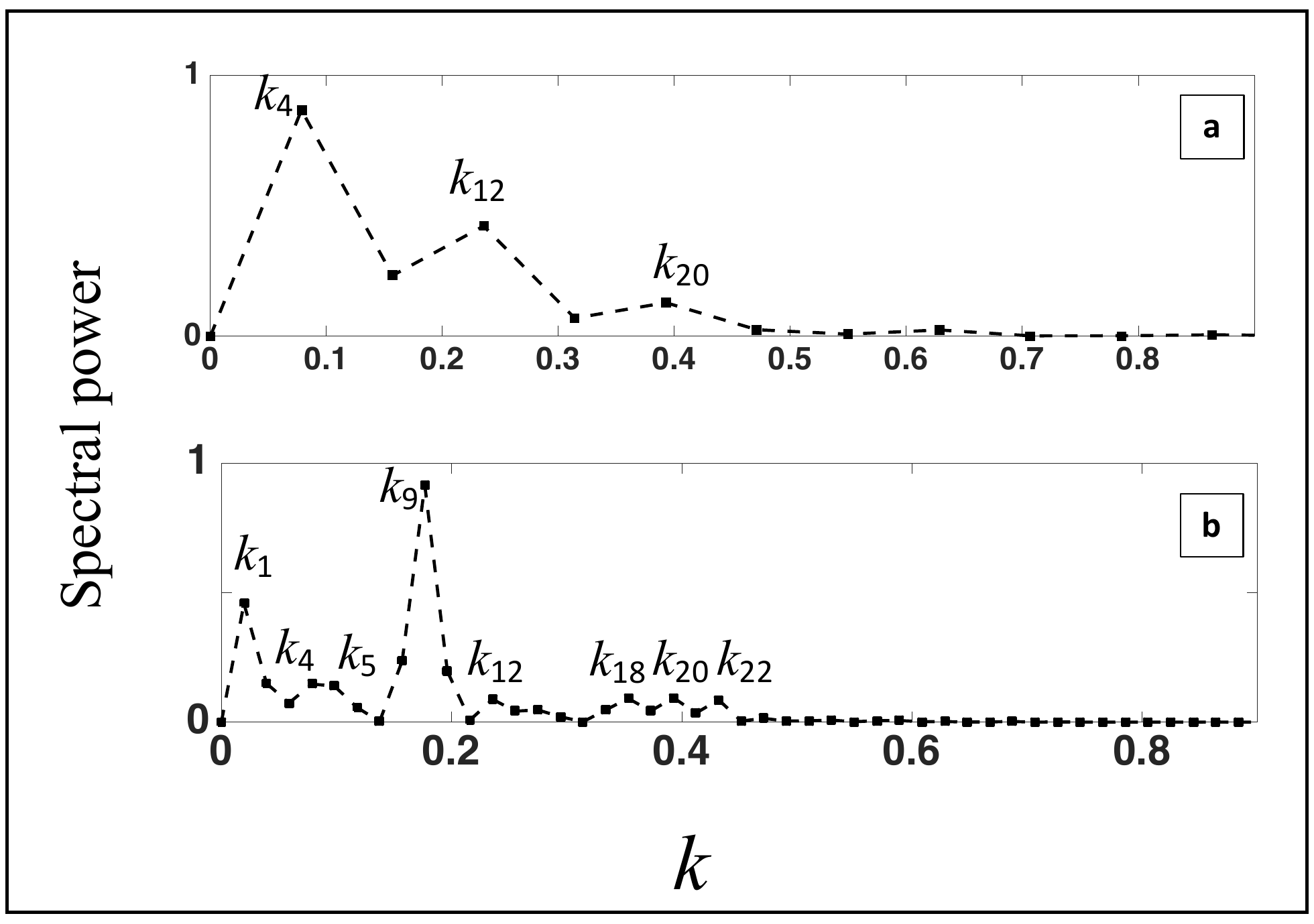}}{\caption{The normalized spectral power of the interface $x=f(y,t)$ for model parameters $\mathrm{MM}(n=1;\mathrm{Le}=0.6;\theta_{\rm{ig}}=0.75)$:  a) $L_{y} =80\, \delta_{\rm{c}}$ at $ t = 76\tau$ , b) $L_{y} =320\, \delta_{\rm{c}}$ at $ t = 69\tau$. Interfaces evolve from a noisy initial condition.  The  modes into which the late-time interface decomposes are indicated for each system size.}\label{fig:2000_Ly320_Ly80_noisy}
  \end{figure}

Figure~\ref{fig:2000_Ly320_Ly80_noisy} presents the spectral power of the 1D interface that develops in the larger and smaller system at late times. These spectrums indicate, as expected, that the front adopts a pattern that is, to lowest order, dominated by the modes available to the system (see Fig.~\ref{fig:dispersion}(b)). The spectral power in Fig.~\ref{fig:2000_Ly320_Ly80_noisy} reveals that the larger system can trigger more of the modes from Fig.~\ref{fig:dispersion}(b) in the states it adopts, consistent with the earlier discussion about the correlation between system size and mode selection. It is found that the peaks comprising the spectral power of the interface rise and fall during its evolution. As anticipated in the quasi-linear regime, the growth rate of each peak during a "rising" phase closely follows the growth rate predicted from the linear dispersion relationship in Fig.~\ref{fig:dispersion}(b)).

\subsubsection{Effect of Lewis number on the morphology}

As discussed earlier, when the value of the Lewis number $\mathrm{Le}$ decreases a rapid transition to strongly non-linear behaviour is expected, signalled by complex interface morphologies and large tip-to-groove amplitudes.   
To further illustrate this, we examined simulations with model parameters $\mathrm{MM}(n=1;\mathrm{Le};\theta_{\mathrm{ig}}=0.75)$ where $0.1 \le \mathrm{Le} \le 0.6$.
Since the time scale $\tau$ in our system depends on the value of the Lewis number (see appendix A), we rescale time for each $\rm {Le}$ studied here according to 

\begin{flalign}
\tau(\rm{Le}^{'}) &= \left(\frac{{1+\frac{\theta_{\rm{ig}}}{\rm{Le}^{'}(1-\theta_{\rm{ig}})}}}{1+\frac{\theta_{\rm{ig}}}{\rm{Le}(1-\theta_{\rm{ig}})}}\right)\,\tau(\rm{Le}),
\end{flalign}
where a reference time-scale, $\tau_{\rm{ref}}$, is chosen to correspond to the largest Lewis number employed ($\rm{Le}=0.6$). 
The size of the computational domains used is given by $L_{x} = 600 \,\delta_{\rm{c}}$ and $L_{y} =160 \,\delta_{\rm{c}}$.
To isolate the effect of $\mathrm{Le}$ we use the same initial conditions in all cases, consisting of a planar steady-state with a sinusoidal perturbation of wavenumber $k=k_{8}=0.05\pi/\delta_{\rm{c}}$.

\begin{figure}[!h]
\centering
\includegraphics[width=0.4\textwidth]{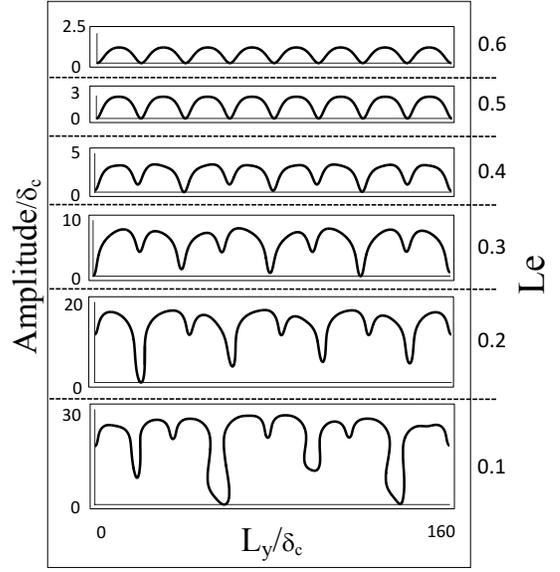}  
\caption{Amplitude of cellular pattern for model parameters $\mathrm{MM}(n=1;\mathrm{Le};\theta_{\mathrm{ig}}=0.75)$ for range of Lewis numbers $0.1\leq\mathrm{Le}\leq0.6$. A sinusoidal perturbation with $k =0.05\pi/\delta_{\rm{c}}$ mode is used for the initial condition. The time of each image is $t_{\rm{Le}=0.1}=1281.33\,\tau_{\rm{ref}}$ , $t_{\rm{Le}=0.2}=885.24\,\tau_{\rm{ref}}$ , $t_{\rm{Le}=0.3}=669.02\,\tau_{\rm{ref}}$,   $t_{\rm{Le}=0.4}=469.24\,\tau_{\rm{ref}}$, $t_{\rm{Le}=0.5}=386.04\,\tau_{\rm{ref}}$, $t_{\rm{Le}=0.6}=332.8\,\tau_{\rm{ref}}$. The reference time $\tau_{\rm{ref}}$ corresponds to $\rm{Le}=0.6$. }
\label{fig:cellular_amp_Lerange}
\end{figure}

Typical late-time interface configurations for different Lewis numbers are shown in Fig.~\ref{fig:cellular_amp_Lerange}. The times indicated in the figure are with respect to the reference time-scale $\tau_{\rm{ref}}$. 
The observed trend is consistent with the results in Fig.~\ref{fig:Linear_weakly_non_linear} and our previous discussion. For the larger values of the Lewis number, the front develops a single mode shallow cellular morphology and reaches a steady-state. With lower values of the Lewis number, highly non-linear effects dominate the evolution of the front, now composed of several modes, non-symmetrical cells of different depths, and overhangs.  

\begin{figure}[!h]
\centering
\includegraphics[width=0.4\textwidth]{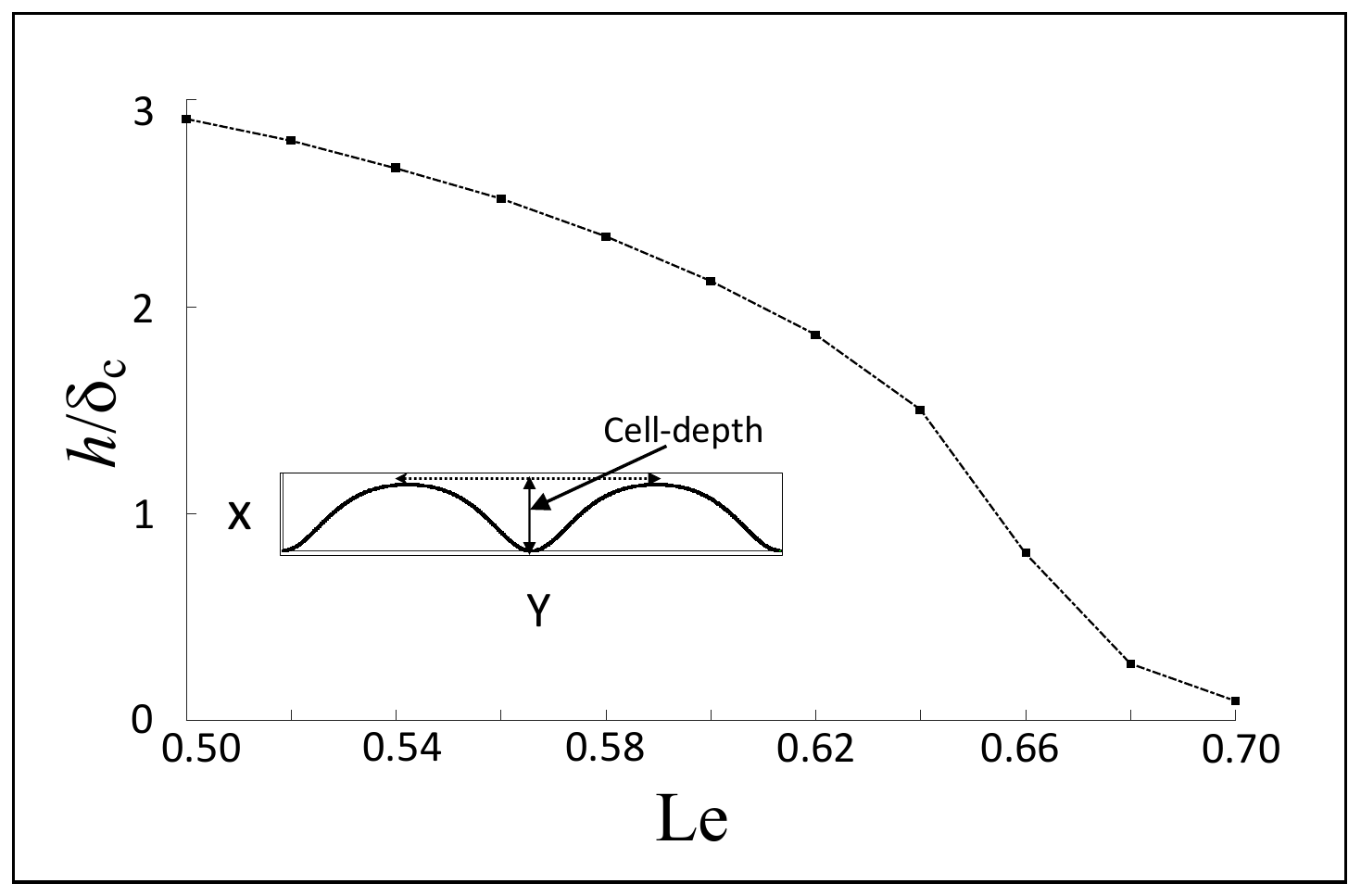}  
\caption{Cell depth parameter $h$ versus ${\rm Le}$ in units of $\delta_{\rm{c}}$. Inset shows  a steady-state cellular interface for model parameters $\mathrm{MM}(n=1;0.5\leq \rm{Le}\leq0.7;\theta_{\mathrm{ig}}=0.75)$ at late time, $t=300\,\tau(\rm{Le})$. Cell depth measures the max-to-min positions on the interface.}
\label{fig:amp_compire_Le}
\end{figure}

Figure~\ref{fig:amp_compire_Le} plots the dependence of the cell amplitude with ${\rm Le}$. Cell depth $h$ is defined as max-to-min positions on the interface. For Lewis number in the range depicted in the figure (\mbox{$0.5\leq\rm{Le}\leq0.7$}) and for the given model parameters, a front develops a single mode steady-state. The average amplitude of the cells decreases with increasing Lewis number and becomes very small as $\rm{Le}$ approaches its critical value ($\rm{Le}_0$). 

\subsection {Non-Linear Front Dynamics for Small $\mathrm{Le}$}

This section examines non-linear combustion fronts in the limit of low Lewis number ($\rm{Le}<0.4$). In this regime, nonlinear effects dominate the dynamics and the front develops complex morphologies involving multiple deep cells with overhangs, or dendritic fingers. 

\begin{figure}[!h]
\centering
\includegraphics[scale = 0.53]{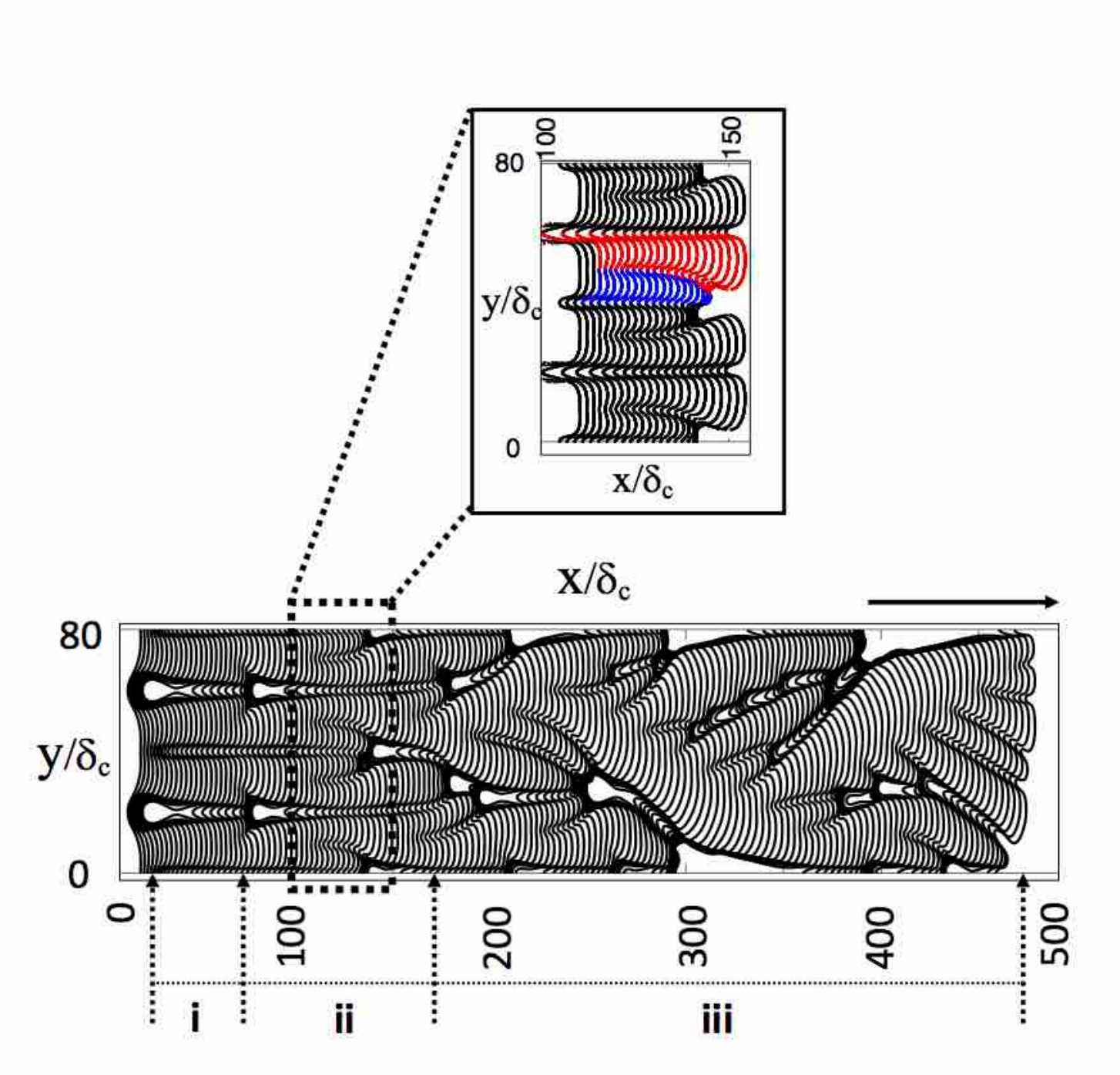}}{\caption{(Colour online) Temporal history of evolution of 1D front corresponding to isotherm $\theta(x,y) =\theta_{\mathrm{ig}}$ for model parameters $\mathrm{MM}(n=1;\mathrm{Le}=0.05;\theta_{\mathrm{ig}}=0.75)$ with initial sinusoidal wave mode $k=0.05\pi/\delta_{\rm{c}}$ and in the system size $L_x=600\,\delta_{\rm{c}} , L_y=80\,\delta_{\rm{c}}$. Dotted arrows define the approximate border between zones \textbf{i}, \textbf{ii}, and \textbf{iii}. Inset: zoom-in of a tip-splitting event and a cell elimination event in region $\bf{ii}$. Colours represent different length scales emerging in the tip-splitting  process, with  \textcolor{red}{red} representing a large cell and \textcolor{blue}{blue} a small cell. The direction of front propagation is shown by the arrow.   }
 \label{fig:fossil_Ly80_Le05}
\end{figure}

We probe the front dynamics for model parameters $\mathrm{MM}(n=1;\rm{Le}=0.05;\theta_{\rm{ig}}=0.75)$.
To investigate the size effect in the non-linear regime of the front evolution, two domain sizes were used, ($L_x =600\,\delta_{\rm{c}}, L_y=80\,\delta_{\rm{c}}$) and ($L_x=600\,\delta_{\rm{c}},L_y=160\,\delta_{\rm{c}}$).
A  planar steady-state solution for each case is perturbed initially with a sinusoidal wave with $k=0.05\pi/\delta_{\rm{c}}$ and  $k=0.025\pi/\delta_{\rm{c}}$, respectively. 
The evolution of the front morphology $x=f(y,t)$ for the smaller domain is shown in Fig.~\ref{fig:fossil_Ly80_Le05}. The combustion front, defined as the isotherm $\theta(x,y)=\theta_{\rm{ig}}=0.75$, is plotted every $\Delta t =2.6\,\tau$. The duration of the simulation was $416\,\tau$. Three stages of the front evolution are indicated in Fig.~\ref{fig:fossil_Ly80_Le05}, regions $\bf{i}$, $\bf{ii}$, and $\bf{iii}$.

At the early stages of stage ${\bf i}$, the perturbed front evolves into crests that have better access to the oxidizer than the troughs since, by geometry, replenishing of burned fuel is more effective ahead of the crests. Thus the crests grow faster than the bottom of the troughs, which remain close to their initial position, and the front develops cells that deepen. When the cells are deep enough they develop a lateral instability. All fingers tip-split, shortly after which elimination of two pairs of new fingers occurs. In Fig.~\ref{fig:fossil_Ly80_Le05}, this corresponds to the end of region $\bf{i}$. As the surviving cells continue to grow and deepen in region $\bf{ii}$, a lateral instability begins to develop again, with some of the tip-split fingers becoming overgrown while others grow until they become unstable themselves at the end of region ${\bf ii}$. Throughout regions ${\bf i}$ and ${\bf ii}$, no steady-state pattern emerges as cells grow, split and compete, with some becoming eliminated while others grow. 

A full-blown non-linear regime follows in region ${\bf iii}$. The front does not establish a dominant growth mode due to system size confinement. Instead, one of the surviving cells from region ${\bf ii}$ undergoes a rapid series of tip-splitting and cell overgrowth events, with one of the tertiary branches eventually overgrowing all previous branches in the system. The evolution of this surviving branch then follows itself a 
series of tip splitting instabilities into primary and tertiary branches. The pattern of side-branching behaviour shown in Fig.~\ref{fig:fossil_Ly80_Le05} is the hallmark of {\it seaweed} dendritic growth.
 
The above process of finger instability, tip-splitting, and cell elimination in region ii are better depicted in the inset of Fig.~\ref{fig:fossil_Ly80_Le05}. It is noteworthy that the mode dominating the front morphology at early stages is fairly close to $k_{\rm{max}}=0.15\pi/\delta_{\rm{c}}$, the fastest growing mode predicted by the linear stability analysis for the same model parameters shown in Fig.~\ref{fig:dispersion}(a). 

\begin{figure}[!h]
\centering
\includegraphics[scale = 0.4]{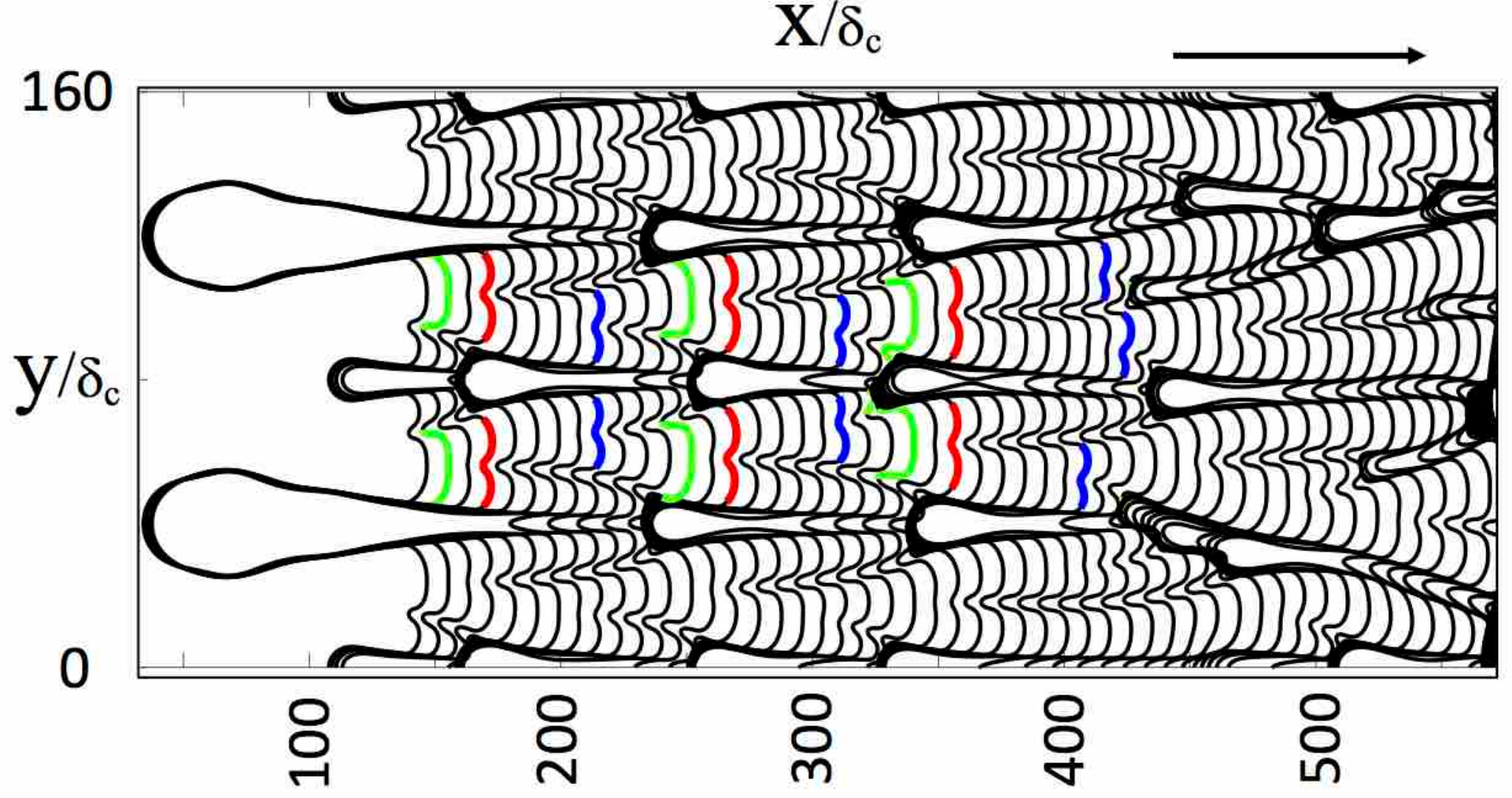}}{\caption{(Colour online) Temporal history of evolution of 1D front corresponds to isotherm $\theta(x,y) =\theta_{\mathrm{ig}}$ for model parameters \mbox{$\mathrm{MM}(n=1;\mathrm{Le}=0.05;\theta_{\mathrm{ig}}=0.75)$}. The initial interface is a sinusoidal wave mode $k=0.025\pi/\delta_{\rm{c}}$ and system size is $L_x=600\,\delta_{\rm{c}} , L_y=160\,\delta_{\rm{c}}$. Colours represents different length scales emerging in the tip-splitting process, with \,\textcolor{green}{green} representing a cell from a primary tip-split event and  \textcolor{red}{red} and \textcolor{blue}{blue} representing cells formed during secondary and tertiary splitting, respectively. The direction of front propagation is shown by the arrow.}
 \label{fig:fossil_Ly160_Le05}
\end{figure}

The evolution of the combustion front for the larger system ($L_y=160 \,\delta_{\rm{c}}$) is shown in Fig.~\ref{fig:fossil_Ly160_Le05}. The isotherm defining the combustion front $\theta(x,y)=\theta_{\rm{ig}}=0.75$ is plotted every $\sim5.12\,\tau$, and the duration of the complete simulation corresponds to $460\,\tau$. The front evolution, qualitatively, follows a similar evolution to that of the smaller system up to the first two stages described above. After the second stage, however, the front  establishes a ``quasi-regular" pattern of growth defined by eight branches (red cells at $x/\delta_c \sim150$). These branches compete and undergo cell splitting, tertiary branching, and at times merging. There are instances where, very briefly, the front goes back to establishing once more four branches (green cells) in the dominant growing pattern, but it quickly splits back to eight branches, which is mostly the state that dominates the pattern from there on in the simulation. 

We compare the numerically approximated $k$ modes associated with various primary (green cells), secondary (red cells), and  tertiary (blue cells) splittings in Fig.~\ref{fig:fossil_Ly160_Le05} with the linear modes from the dispersion relation (Fig.~\ref{fig:dispersion}(a)) for the given model parameters ($\rm{Le =0.05} , \theta_{\rm{ig}}=0.75$) and the corresponding system size. Our analysis shows that the length scale of the green cells corresponds to an approximate wavenumber $k=k_{11}=0.1375\pi/\delta_{\rm{c}}$, which  is remarkably close to \mbox{$k_{\rm{max}} =0.15\pi/\delta_{\rm{c}}$}.  These cells, however, split into the eight branches (red cells) with dominant mode $k=k_{8}=0.1\pi/\delta_{\rm{c}}$, which is not close to the fastest growing mode in the corresponding dispersion relation. A detailed analysis of spacing selection in the non-linear regime is beyond the scope of this work. Spacing selection in the non-linear regime has a long history in solidification studies \cite{Amoorezaei2010,Gurevich2010a} and while insight can be gained from analyses like the one above, it still represents a largely unsolved problem.  

The difference in the front evolution in the two cases examined in this section can be tracked to the difference in system width. Initially both systems develop four main branches with approximate wave mode  $k=0.15\pi/\delta_{\rm{c}}$. The larger system then transitions to establishing eight branches with a wavenumber approximately $k=0.1\pi/\delta_{\rm{c}}$, while the smaller system fails to do so due to system size effects that hinder the establishment of a recurring pattern to the complex interface morphology.  

\section{Conclusions}
\label{conc}
We introduced a unified mathematical model, so-called  ``master model", for solid fuel combustion in the presence of an oxidizer for zero and first-order kinetics, with different heterogeneity of reactant. 
The gas-less limit ($\rm{Le} \to \infty$), for both kinetic orders ($n=0,1$), was examined and connected to previously established models \cite{sivashinsky1981spinning,matkowsky1978propagation}. 

We found that the combustion front develops a cellular morphology below a critical value of the Lewis number. The linear stability of these cellular fronts was investigated numerically and found to agree very well with the dispersion relation obtained analytically by Brailovsky et al.
We addressed the influence of the size of the computational domain on mode selection, and explored how the front dynamics varies with the value of the Lewis number.
Our results indicate that lowering the value of the Lewis number broadens the range of instability modes that can be triggered and increases the growth rates of the most unstable $k$ modes.

For values of the Lewis number $\rm{Le}$ near the critical value, the transition from the linear to the non-linear (i.e, late-time) regime becomes prolonged, and we identified this regime as a ``quasi-linear" regime. The dynamics of this regime were explored and discussed.
For lower Lewis number values \mbox{($\rm{Le}<0.4$)} the transition to the fully non-linear regime occurs early in the evolution of the front. Non-linear effects dominate the dynamics of the front and large amplitude cellular structures develop. We briefly explored the non-linear regime and the influence of finite size effects on the late-time dynamics of the front and its morphology, emphasizing the similarities of the combustion front in this regime with isotropic dendritic microstructures that have been reported in solidification modelling.

Our examination of combustion fronts will continue in an upcoming publication, which will address the front dynamics considering discrete fuel particles rather than a continuum fuel source, as well as stochasticity.

\section{Acknowledgments}
This work was supported by a Canadian Space Agency Class Grant ``Percolating Reactive Waves in Particulate Suspensions''. We thank Compute Canada for computing resources.\\

\section{Appendix A: \textit{Master model} derivation}
The coupled Reaction-diffusion equations describing the combustion of solid fuel with ignition temperature kinetics, and neglecting solid fuel diffusion, can be written as: 

\begin{flalign}
\left\{
\begin{array}{ll}
\label{eq:RD-eqs1}
\rho\,c_{p}\,\frac{\partial{{T(\vec{x})}}}{\partial{t}}= \kappa\,\nabla^2{T(\vec{x})} + \Lambda\,W  \\
&\\
\frac{\partial{C(\vec{x})}}{\partial{t}}=-W, 
\end{array}
\right.
\end{flalign}

\noindent where the source term is given by:

\begin{flalign}
\label{eq:RD-rateterm}
W &= \frac{1}{t_{\rm{r}}}\,\mathrm{H}(T-T_{\mathrm{ig}})\, C(\vec{x}), 
\end{flalign}

\noindent while $\Lambda$, $C(\vec{x})$, $t_{\rm{r}}$ and $T_{\mathrm{ig}}$ represent the heat of reaction [J/kg], the local solid fuel concentration [kg/$\mathrm{m^{3}}$], the reaction time [s] and the ignition temperature [K], respectively.

In Eq.~(\ref{eq:RD-rateterm}), the condition of source activation above $T_{\rm{ig}}$ appears in the form of a Heaviside function in temperature. 
Normalizing the local concentration with respect to the initial solid fuel concentration $B$[kg/$\mathrm{m^{3}}$], $\phi_{\mathrm{s}}(\vec{x})=C(\vec{x})/B$, Eqs.~(\ref{eq:RD-eqs1})-(\ref{eq:RD-rateterm}) become:

\begin{flalign}
\left\{
\begin{array}{ll}
\label{eq:RD-eqs2}
\rho\,c_{p}\,\frac{\partial{{T(\vec{x})}}}{\partial{t}} = \kappa\,\nabla^2{T(\vec{x})} + \Lambda\,\frac{B\,\phi_{\mathrm{s}}(\vec{x})\,\mathrm{H}(T-T_{\mathrm{ig}})}{t_{\rm{r}}} \\ 
& \\
\frac{\partial{\phi_{\mathrm{s}}(\vec{x})}}{\partial{t}} =-\frac{\phi_{\mathrm{s}}(\vec{x})\,\mathrm{H}(T-T_{\mathrm{ig}})}{t_{\rm{r}}},         
\end{array}
\right.
\end{flalign}

By converting the local concentration of solid fuel, $\phi_{\mathrm{s}}(\vec{x})$, to the corresponding local concentration of oxidizer (deficient reactant) $\phi(\vec{x})$, and taking into account the chemical stoichiometric ratio $\gamma$, Eqs.~(\ref{eq:RD-eqs2}) then leads to

\begin{flalign}
\left\{
\begin{array}{ll}
\label{eq:RD-eqs3}
\frac{\partial{{T(\vec{x})}}}{\partial{t}} =  \alpha\,\nabla^2{T(\vec{x})} + q\,\frac{\phi(\vec{x})\,\mathrm{H}(T-T_{\mathrm{ig}})}{t_{\rm{r}}}  \hspace{24pt} \ \\ 
& \\
\frac{\partial{\phi(\vec{x})}}{\partial{t}}=  D\,\nabla^2\mathrm{\phi}-\frac{\phi(\vec{x})\,\mathrm{H}(T-T_{\mathrm{ig}})}{t_{\rm{r}}},         
\end{array}
\right.
\end{flalign}

 \noindent where $\mathrm{q} =\gamma\Lambda B/\rho\,c_{p}$ is the heat of reaction 
 in the same units as the temperature
and is equivalent to the difference between adiabatic and initial temperature of the mixture ($T_{\mathrm{ad}} - T_{\mathrm{{o}}}$), while $\alpha =\kappa/\rho\,c_{p}$ and $D$ are the thermal diffusivity and the oxidizer mass diffusivity, respectively. 

By transforming equations~(\ref{eq:RD-eqs3}) to the co-moving coordinate $\xi =x-ut$, where $u$ is the velocity of a steady-state planar front, and using the far-field values 

\begin{align}
\label{eq:far-field-appdix}
T(+\infty ) =  T_{\mathrm{o}},\hspace{20pt}\ T(-\infty ) = T_{\rm{ad}}, &  \\
 \phi(+\infty ) =  \phi_{\mathrm{o}},\hspace{33pt}\phi(-\infty)  = 0, \nonumber
\end{align}

\noindent along with the continuity of the fields and their derivatives across the interface ($\xi =0$), it is straightforward to derive the steady-state solutions for the temperature,
\begin{flalign}
T_{\rm{ss}}(\xi) =
\left\{
\begin{array}{ll}
\label{eq:SS_temp}
\frac{-1}{\alpha t_{\rm{r}} u^{2}(1+\eta)(\frac{\eta}{D})^{2}+\frac{\eta}{D}}\exp[\frac{u(-1+\eta)}{D}\xi] \hspace{12pt} \xi \leq0 \\   
  & \\                    
(T_{\rm{ig}}-T_{\mathrm{o}})\exp[-\frac{u}{\alpha}\xi] +T_{\mathrm{o}} \hspace{39pt} \xi >0,
\end{array}
\right.
\end{flalign}
and for the concentration,
\begin{flalign}
\phi_{\rm{ss}}(\xi) =
\left\{
\begin{array}{ll}
\label{eq:SS_conc}
\frac{1}{1+\eta}\exp[\frac{u\eta}{D}\xi] \hspace{84pt} \xi \leq0 \\   
  & \\                    
\frac{-\eta}{1+\eta}\exp[-\frac{-u}{D}\xi]+1\hspace{64pt} \xi >0,
\end{array}
\right.
\end{flalign}
where 
\begin{align}
\label{eq:eta}
\eta &=\frac{-1+\sqrt{1+\frac{4D}{u^2t_{\rm{r}}}}}{2}, \hspace{10pt}
\end{align}
Using the continuity condition for temperature across the interface, and introducing a dimensionless temperature 
\mbox{$\theta=(T-T_{\mathrm{o}})/(T_{\rm{ad}}-T_{\mathrm{o}}$)},  gives the steady-state velocity as
\begin{flalign}
\label{eq:SS-velocity_dimless}
u &=\sqrt{\frac{\alpha}{t_{\rm{r}}\big(\frac{\theta_{\rm{ig}}}{1-\theta_{\rm{ig}}}\big)(1+\frac{1}{\mathrm{Le}} \big(\frac{\theta_{\rm{ig}}}{1-\theta_{\rm{ig}}}\big)}}, 
\end{flalign}
Defining the characteristic length and time as
\begin{align}
\delta_{\rm{c}}=\frac{\alpha}{u}\,\,\,\,\,\,, \hspace{10pt}\tau=\frac{\alpha}{u^{2}},
\end{align}
leads to the dimensionless form of Eqs.~(\ref{eq:RD-eqs3}) given by  
\begin{flalign}
\left\{
\begin{array}{ll}
\label{eq:RD-eqs4}
\frac{\partial{\theta(\vec{x})}}{\partial{t}} &=  \alpha\nabla^2\theta(\vec{x}) + \frac{\tau}{t_{\rm{r}}}\,\phi(\vec{x})\,\mathrm{H}(\theta-\theta_{\mathrm{ig}})  \\ 
& \\
\frac{\partial{\phi(\vec{x})}}{\partial{t}} &=  \frac{1}{\mathrm{Le}}\nabla^2\mathrm{\phi}- \frac{\tau}{t_{\rm{r}}}\,\phi(\vec{x})\,\mathrm{H}(\theta-\theta_{\mathrm{ig}})
\end{array}
\right.
\end{flalign}
Introducing $A = \tau/t_{\rm{r}}$, Eqs.~(\ref{eq:RD-eqs4}) take the final form,
\begin{flalign}
\left\{
\begin{array}{ll}
\frac{\partial{\theta(\vec{x})}}{\partial{t}} &= \nabla^2\theta(\vec{x}) + W \hspace{35pt} \ \\   
  & \\                     
\frac{\partial{\phi(\vec{x})}}{\partial{t}} &= \frac{1}{\rm{Le}}\nabla^2\phi(\vec{x}) - W,     
\end{array}
\right.
\end{flalign}
with the source term 
\begin{flalign}
W= 
\left\{
\begin{array}{ll}
 {A\,\phi(\vec{x})\,\mathrm{H}(\theta-\theta_{\mathrm{ig}})}  \hspace{25pt} \mathrm{for}  \  n = 1 \\ 
& \\
 {A\,\mathrm{H}(\theta-\theta_{\mathrm{ig}})\,\mathrm{H}(\phi) } \hspace{24pt}    \mathrm{for}  \  n = 0  \\
\end{array}
\right.
\end{flalign}


%

\end{document}